\documentclass[usenatbib,usegraphicx]{mn2e}
\usepackage{graphicx}
\usepackage{color}
\usepackage{amssymb}
\usepackage{natbib}

     \def\ma{$M_{\rm 2}$}

\def\vs40{\vskip 40pt}        \parskip=0pt
    
\def\kms{km\,s$^{-1}$}

\def\DV{$\Delta V$}  
\def\b-v{($B-V$)\hs2} \def\u-b{($U-B$)\hs2}
\def\B-V{($B-V$)} \def\U-B{($U-B$)}

   \def\vs#1{\vskip#1pt}
 \def\hs#1{\hskip#1pt}

\def\Msun{M$_{\odot}$}         \def\Rsun{R$_{\odot}$}
\def\Lsun{L$_{\odot}$}
           
          \def\deg{\hbox{$^{\circ}$\hskip-3pt .}}
          
\def\K1{{\it K}\hskip-1pt$_1$}      \def\K2{{\it K}\hskip-1pt$_2$}
   
\def\a2sini{$a_2$\hs2 sin\hs2$i$}   
\def\m2sin3i{$m_2$\hs2 sin$^3$\hs2 $i$}
\def\m1sin3i{$m_1$\hs2 sin$^3$\hs2 $i$}
  \def\mags{\raise.8ex\hbox{\srm m}\hs{-1}}
\def\fracp#1/#2{\leavevmode\kern.05em\raise.6ex\hbox{\the\scriptfont0
    #1}\kern-.15em/\kern-.15em\lower.3ex\hbox{\the\scriptfont0 #2}\hs1}
          
\def\V12{$V_{12}$} \def\DV{$\Delta V$} \def\AV{$A_V$}
\def\Ta{$T_{\rm 1}$}  \def\Tb{$T_{\rm 2}$}

\font\smtt=cmtt10 at 5pt  
\def\hs#1{\hskip#1pt}
\def\H2{I\hs{-1.5}\raise0.3ex\hbox{-\hs{-2.0}-}\hs{-2.0}I\hs{-5.5}\lower.05ex
    \hbox{{\smtt\char'174}\hs{-3.25}\lower.42ex\hbox{$_{\circ}$}}\hs4}

\def\dOS{\Delta_{\rm OS}}
\def\tl{{\raise 0.3ex\hbox{$\sc  {\th < \th }$}}}

\title[Models for Giant Binaries]{Models for Sixty Double-Lined Binaries containing Giants}

\author[P. P. Eggleton \& K. Yakut]
       {Peter P. Eggleton$^{1}$\thanks{E-mail: peter.eggleton@yahoo.com}
        Kadri Yakut$^{2}$\\
$^{1}$ Lawrence Livermore National Laboratory, 7000 East Ave, Livermore, CA94551, USA\\
$^{2}$ Department of Astronomy \& Space Sciences, University of Ege, 35100, Bornova--{\.I}zmir, Turkey\\
}
\date{Accepted XXX. Received YYY; in original form ZZZ}

\pubyear{2016}
\begin{document}
\date{Received}
\maketitle
\label{firstpage}
\begin{abstract}
The observed masses, radii and temperatures of 60 medium- to long-period
binaries, most of which contain a cool, evolved star and a hotter less-evolved
one, are compared with theoretical models which include (a) core convective
overshooting, (b)~mass loss, possibly driven by dynamo action as in RS~CVn
binaries, and (c)~tidal friction, including its effect on orbital period through
magnetic braking. A reasonable fit is found in about 42 cases, but in 11 other
cases the primaries appear to have lost either more mass or less mass than the
models predict, and in 4 others the orbit is predicted to be either more or less
circular than observed.  Of the remaining 3 systems, two ($\gamma$~Per and HR~8242)
have a markedly `over-evolved' secondary, our explanation being that
the primary component is the merged remnant of a former short-period {\it
sub}-binary in a former triple system. The last system (V695~Cyg)
defies any agreement at present.
\par Mention is also made of three other systems (V643~Ori, OW~Gem and V453~Cep),
which are relevant to our discussion.
\end{abstract}


\section{Introduction}

It has long been recognized that analyses of binary stars yield far more
precise information regarding stellar age and evolutionary status than can be
derived for single stars, and to that end numerous studies have been made of
double-lined binaries, mostly of short-period eclipsing double-lined
main-sequence (ESB2) systems.  As seen in the review by
\citet{Torres.Andersen.2010}, many can present masses with claimed precisions
of the order of 3\% or better.  The studies by (in particular)
\citet{Demarque.al.1994}, \citet{Claret.1995}, \citet{Pols.al.1997},
\citet{Girardi.al.2000}, \citet{Ribas.al.2000}, \citet{Young.al.2001} and
\citet{Claret.2004} generally show a reasonable agreement with theoretical
models of stellar evolution, although the concept of core convective overshooting
had to be introduced \citep{Maeder.1975, Andersen.1991} in order to account for
a substantially broader main-sequence band than the one that was indicated by
models that did not include overshooting.  But while double-lined main-sequence
binaries provide important constraints on theoretical models \citep[as
demonstrated, for example, by][]{Pols.al.1997}, the constraints on stellar
evolution theory which can be derived from binaries with a post-main-sequence
component -- particularly if one component is evolved to a cool giant and the
other is markedly less evolved -- can be substantially tighter, despite the
fact that the precision of the masses can be more like 10\% than 3\%.  One such
study was made by \citet{Schroder.al.1997}, and this paper builds on it and
extends their sample of 9 systems to 60.

Binaries which contain an evolved component are usually more widely separated
than main-sequence ones, and most do not eclipse.  The great majority of the
binaries in our sample consist of a cool (G--K) giant plus a hot (B--A)
main-sequence companion.  Measured physical parameters for them have
been taken from the literature.
Several of the systems were formally classified as
`Composite-Spectrum Binaries' in the {\it Henry Draper Catalogue}, where most
of them were assigned two HD numbers.

In all of the cases considered here, there is a well-determined spectroscopic
orbit for the evolved star; some have astrometric orbits as well. In
principle, therefore, in order to derive the system's mass ratio it should only
be necessary to measure the radial velocity (RV) of the companion {\it once},
at a favorable quadrature phase whose dates can be calculated from the
spectroscopic orbit of the primary. But in a surprising number of cases -- at least
6 out of 46, or 13\% -- it is found that the hot companion is itself a component of
a short-period {\it sub}-binary (R. E. M. Griffin, p.c.).  Many RV measurements of all systems at
different phases are therefore necessary, either to eliminate the possibility
of a third body or to determine the sub-orbit.  Moreover, one result of the
present paper is to suggest that two cases out of the 60 are best understood as
{\it former} triples but which are now binaries because the inner pair merged.

In addition to the problem of possible sub-binarity, there are many practical
reasons why the analysis of a composite spectrum is more troublesome than for
shorter-period ESB2s.  As \citet{GriffinRR.1986} describes, the attainable
accuracy depends on the {\it nature} of the secondary's spectrum as well as on
methods of isolating and measuring it, and when the lines available for RV
measurement are few (as in early A-type dwarfs) and those that are available
are also broadened by rapid rotation (as often happens), the precision of the
measured mass ratio of that system will be rather limited.  Nevertheless, even
the more ragged ones can still provide a very useful check on theoretical
evolutionary models.

Of the systems that prove to be triple, it usually happens that the hotter
component consists of a shorter-period sub-binary whose members are either two
similar-mass main-sequence stars (in which case the system is triple-lined) or
a main-sequence star plus a cooler, fainter dwarf (in which case only the two
brighter spectra are visible but the presence of the third star is revealed by
RV vagaries of large amplitude in the secondary's spectra).  Quite often,
therefore, a triple system may initially contain 3 components of fairly
comparable mass. If the most massive of the three is itself in a close
sub-binary with the least massive, one can formulate an evolutionary path for
the close pair that leads to a merger, as recently observed in V1309~Sco
\citep{Tylenda.al.2011}. That may then explain how a system can have a
secondary which is conspicuously less massive than its primary, yet is evolved
some considerable way across the main sequence band -- as seems to be true of
two systems in our sample.

In the last decade many ground and space based photometric surveys
(e.g. OGLE, ASAS, CoRoT, Kepler, Gaia) provided accurate light variations
from both single and binary stars.
The combination of highly sensitive photometric data with ground-based spectroscopic data
leads to very accurate orbital and physical parameters of binary systems.
Hence, this helps us to test current stellar evolution theories in a more sensitive way.
In this study, we use an important amount of systems observed with these projects.
\S2 presents the basic principles that have been adopted for modelling the
systems, and gives examples of the agreement (or otherwise) with observation.
The models of overshooting, tidal friction and stellar wind are discussed
in \S3.1, \S3.2 and \S3.3, respectively, and the results are described on a
case-by-case basis in \S4. An algorithm for assessing the `goodness of fit'
between observed and theoretical models is briefly described in \S4.1, and
more extensively in Appendices B and C. Two possible former triples are
described in \S5.1, while a system that presently defies a tenable explanation
is discussed in \S5.2. Our conclusions are summarized in \S6.  The quality of
the agreements between model and observation is best assessed graphically, as
shown for 10 systems in Figs.~1--4; all 60 Figures are available online.

\section{Adopted principles for selecting and modelling the sample}

The 60 binary systems discussed in this paper are listed in
Table~1, where a number of aliases, and the primary literature
references, are also listed.

\setcounter{table}{0}
\begin{table*}
\setlength{\tabcolsep}{0.050in}
\caption{Aliases and Basic References for the Sample of 60 Systems
\label{tab:1new}}
\indent  \\
\begin{tabular}[]{ccll|l}
\hline
No. &Short name used here   & One or more conventional IDs                        & Principal References        \\
\hline
 1  & SMC-130               & OGLE SMC130.5 4296, 2MASS J00334789-7304280         & Graczyk et al.~(2014)        \\
 2  & SMC-126               & OGLE J004402.68-725422.5, 2MASS J00440266-7254231   & Graczyk et al.~(2014)         \\
 3  & SMC-101               & OGLE SMC130.5 4296, 2MASS J00334789-7304280         & Graczyk et al.~(2014)          \\
 4  & HD 4615               & HD 4615/6, HIP 3787                                 & Griffin \& Griffin (1999)       \\
 5  & $\eta$ And            & HR 271, HD 5516, HIP 4463, SBC9-50                  & Schr\"oder et al.~(1997)         \\
 6  & SMC-108               & OGLE SMC-SC8 201484, 2MASS J01001803-7224078        & Graczyk et al.~(2013)              \\
 7  & BE Psc                & HD 6286, HIP 5007, SBC9-2802                        & Strassmeier et al.~(2008)           \\
 8  & AS-010538             & ASAS J010538 --8003.7                               & Ratajczak et al.~(2013)              \\
 9  & AI Phe                & HD 6980, HIP 5438, SBC9-61                          & Andersen et al.~(1988)                \\
10  & $\tau$ Per            & HR 854, HD 17878/9, HIP 13531, SBC9-148             & Griffin et al.~(1992); Ake \& Griffin (2015)\\
11  & $\gamma$ Per          & HR 915, HD 18925/6, HIP 14328, SBC9-154             & Griffin (2007)                \\
12  & TZ For                & HD 20301, HIP 15092                                 & Andersen et al.~(1991)         \\
13  & HR 1129               & HD 23089/90, HIP 17587                              & Griffin et al.~(2006)           \\
14  & OGLE-Cep              & OGLE LMC--CEP --227                                 & Pilecki et al.~(2013)            \\
15  & RZ Eri                & HD 30050, HIP 22000, SBC9-270                       & Popper (1988)                     \\
16  & OGLE-01866            & OGLE LMC-ECL-1866, MACHO 47.1884.17                 & Pietrzy{\'n}ski et al.~(2013)         \\
17  & OGLE-03160            & OGLE LMC-ECL-03160, MACHO 18.2475.67                & Pietrzy{\'n}ski et al.~(2013)          \\
18  & $\zeta$ Aur           & HR 1612, HD 32068/9, HIP 23453, SBC9-292            & Griffin (2005); Ake \& Griffin (2015)  \\
19  & OGLE-06575            & OGLE LMC-ECL-06575,  MACHO 1.3926.29                & Pietrzy{\'n}ski et al.~(2013) \\
20  & OGLE-EB               & OGLE J051019.64 --685812.3, OGLE LMC-ECL-9114       & Pietrzy{\'n}ski et al.~(2009) \\
21  & OGLE-09660            & OGLE LMC-ECL-09660, MACHO 52.5169.24                & Pietrzy{\'n}ski et al.~(2013)    \\
22  & OGLE-10567            & OGLE LMC-ECL-10567, MACHO 2.5509.50                 & Pietrzy{\'n}ski et al.~(2013)     \\
23  & OGLE-26122            & OGLE LMC-ECL-26122, MACHO 79.5500.60                & Pietrzy{\'n}ski et al.~(2013)  \\
24  & $\alpha$ Aur          & HR 1708, HD 34029, HIP 24608, SBC9-306              & Weber \& Strassmeier (2011)    \\
25  & OGLE-15260            & OGLE LMC-ECL-15260, MACHO 77.7311.102               & Pietrzy{\'n}ski et al.~(2013)    \\
26  & $\delta$ Ori          & HR 1852, HD 36486, HIP 25930                        & Richardson et al.~(2015)   		  \\
27  & HR 2030               & HD 39286, HIP 27747                                 & Griffin \& Griffin (2000b) 		   \\
28  & V415 Car              & HR 2554, HD 50337, HIP 32761, SBC9-424              & Komonjinda et al.~(2011)    		\\
29  & HR 3222               & HD 68461, HIP 40231                                 & Griffin \& Griffin (2010)    		 \\
30  & AL Vel                & HIP 41784, SBC9-519                                 & Kilkenny et al.~(1995); Eaton (1994)  \\
31  & RU Cnc                & HIP 42303, SBC9-525                                 & Imbert (2002)             \\
32  & 45 Cnc                & HR 3450, HD 74228,  HIP 42795                       & Griffin \& Griffin (2015)  \\
33  & $o$ Leo               & HR 3852, HD 83808/9, HIP 47508, SBC9-580            & Griffin (2002)               \\
34  & DQ Leo                & HR 4527, HD 102509, 93 Leo, HIP 57565, SBC9-690     & Griffin \& Griffin (2004)   \\
35  & 12 Com                & HR 4707, HD 107700, HIP 60351, SBC9-719             & Griffin \& Griffin (2011)     \\
36  & 3 Boo                 & HR 5182, HD 120064, HIP 67239, SBC9-780             & Holmberg et al.~(2009)          \\
37  & HR 5983               & HD 144208, HIP 78649, SBC9-880                      & Griffin \& Griffin~(2000a)       \\
38  & HR 6046               & HD 145849, HIP 79358, SBC9-892                      & Scarfe et al.~(2007)              \\
39  & AS-180057             & ASAS J180057-2333.8, TYC 6842-1399-1                & Suchomska et al.~(2015)             \\
40  & AS-182510             & ASAS J182510 --2435.5                               & Ratajczak et al.~(2013)             \\
41  & V1980 Sgr             & HD 315626, ASAS J182525-2510.7                      & Ratajczak et al.~(2013)               \\
42  & V2291 Oph             & HR 6902, HD 169689/90, HIP 90313, SBC9-1050         & Griffin et al.~(1995)                  \\
43  & 113 Her               & HR 7133, HD 175492, HIP 92818, SBC9-1100            & Parsons \& Ake (1998); Pourbaix \& Boffin (2003)\\
44  & KIC~10001167          & 2MASS J19074937+4656118, TYC 3546-941-1             & Rawls (2016); He{\l}miniak et al.~(2016)\\
45  & KIC~5786154           & 2MASS J19210141+4101049                             & Rawls (2016)                            \\
46  & KIC~3955867           & 2MASS J19274322+3904194                             & Rawls (2016)                              \\
47  & KIC~7037405           & 2MASS J19315429+4232516                             & Rawls (2016)                               \\
48  & 9 Cyg                 & HR 7441, HD 184759/60, HIP 96302                    & Griffin et al.~(1994)                        \\
49  & SU Cyg                & HR 7518, HD 186688, HIP 97150, SBC9-2142            & Evans \& Bolton (1990)                        \\
50  & $\delta$ Sge          & HR 7536, HD 187076, HIP 97365, SBC9-1174            & Schr\"oder et al.~(1997); Griffin (1991)      \\
51  & V380 Cyg              & HR 7567, HD 187879, HIP 97634, SBC9-1180            & Pavlovski et al.~(2009)        \\
52  & HD 187669             & ASAS J195222-3233.7, 2MASS J19522207-3233396        & He{\l}miniak et al.~(2015)       \\
53  & HD 190585             & KIC 9246715, BD+45 3047                             & Rawls (2016)                     \\
54  & HD 190361             & HIP 98791                                           & Griffin \& Griffin (1997)         \\
55  & V695 Cyg              & 31 Cyg, HR 7735, HD 192577, HIP 99675, SBC9-1215    & Griffin (2008)                     \\
56  & V1488 Cyg             & 32 Cyg, HR 7751, HD 192909/10, HIP 99848, SBC9-1218 & Griffin (2008)                       \\
57  & QS Vul                & 22 Vul, HR 7741, HD 192713, HIP 99853, SBC9-1216    & Eaton \& Shaw (2007); Ake \& Griffin (2015)\\
58  & $\alpha$ Equ          & HR 8131, HD 202447/8, HIP 104987, SBC9-1291         & Griffin \& Griffin (2002)                   \\
59  & HR 8242               & HD 205114/5, HIP 106267, SBC9-1312                  & Burki \& Mayor (1983)                        \\
60  & HD 208253             & HIP 108039                                          & Griffin \& Griffin (2013)                     \\
 \hline
\end{tabular}
\end{table*}

\begin{figure*}  
\centering
\begin{tabular}[]{lll}
\hskip-10pt
\includegraphics[height=6.0cm]{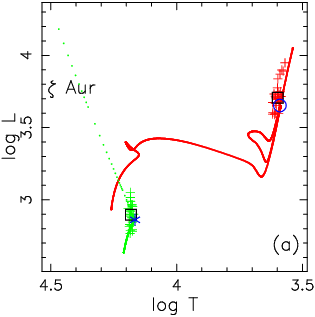} &
\hskip-10pt
\includegraphics[height=6.0cm]{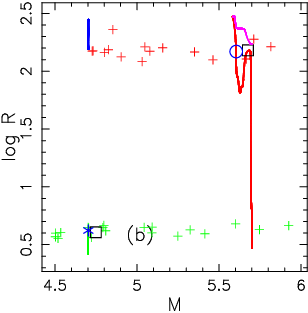} &
\hskip-10pt
\includegraphics[height=6.0cm]{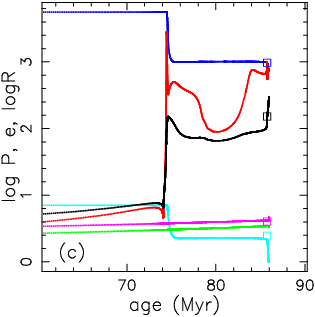} \\
\hskip-10pt
\includegraphics[height=6.0cm]{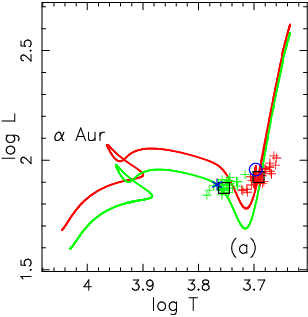} &
\hskip-10pt
\includegraphics[height=6.0cm]{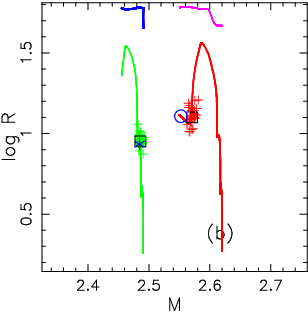} &
\hskip-10pt
\includegraphics[height=6.0cm]{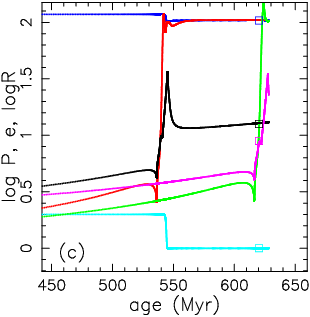} \\
\end{tabular}
\caption{Evolutionary tracks for the components of $\zeta$~Aur (upper) and
$\alpha$~Aur (lower).  Panel (a) shows the $(\log L,\ \log T)$ plane.  Observed
values are plotted as squares.  A blue circle on the primary's track (red)
indicates a place where the model agrees reasonably well with the data, and a
blue asterisk on the secondary's track (green) is the co\ae val point. Plusses
(red or green) indicate a random distribution of errors according to a normal
distribution and the published values of $\sigma$ (Table~3). If the scatter is large,
several of the 30 randomly-varied plusses may be absent from a panel.  Panel
(b) shows the (log\,$R$, $M$) plane; the corresponding Roche-lobe radii are
indicated in dark blue and light blue near the top of each plot.  Panel (c)
shows the time-evolution of orbital period (dark blue), both spin periods (red,
green), eccentricity (pale blue), and both radii (black, purple). Only the last
third of the evolutionary time is plotted. Observed values of period, radii and
eccentricity are shown as small squares, without the scatter.}
\label{fig:fig1}
\end{figure*}

The evolutionary code developed and used here solves for both stars {\it
simultaneously} (Yakut \& Eggleton 2005), including orbit and spin; however,
near-uniform rotation is assumed for each component, as recommended by
\citet{Spruit.1998}.  Tidal friction is incorporated, so that spin
period, orbital period and orbital eccentricity are allowed to modify each
other.  Also included is a model of dynamo-driven winds, such as are expected
in RS~CVn binaries \citep{Biermann.Hall.1976} and also in BY~Dra binaries
\citep{Bopp.Evans.1973}.  Combining tidal friction and dynamo-driven wind means
that magnetic braking affects not just the component spins but also their
orbital periods. The code also contains a necessarily rather crude model of
core convective overshooting, which is quite considerably constrained by
comparing the models with some of the observed systems.

In their review of ESB2 systems \citet{Torres.Andersen.2010} listed 95 ESB2 binaries
for which they concluded that the masses and radii are precise to better than
3\%. However, only three of the 190 components in that sample are red giants;
two are in a remarkable eclipsing binary in the LMC \citep[OGLE 051019;][]{Pietrzynski.al.2009}
referred to here as OGLE-EB, and one is the primary of TZ~For. A third
system (AI~Phe) has a K0\,IV subgiant that is well beyond the main sequence,
but is still only near the bottom of the first giant branch. In fact that
sample contains several other components classified spectroscopically as
subgiants and even giants, but they are apparently still within the
main-sequence band. Torres et al.~also listed 23 astrometric spectroscopic
binaries whose component masses were known with similar precision; one
($\alpha$~Aur) has two giant components, although the secondary is actually in
the Hertzsprung gap rather than on the first giant branch, and another (o~Leo)
has a primary that is also clearly in the Hertzsprung gap. These five systems
are included in our set of 60.

Finding a good fit between a theoretical binary and an observed one belonging
to the category studied here is considerably more tricky than for double
main-sequence binaries, for a number of reasons.  The main one is a major
non-linearity, since the star and its model may have the same radii at three or
even five different points in its evolution. A model has a short-lived local maximum
followed by a local minimum at the terminal main sequence; it may have
another local maximum and  minimum near the base of the first giant branch before growing
substantially until core-He ignition. It then reaches a long-lived local
minimum radius during the GK-giant clump stage, and increases again towards the
second or asymptotic giant branch, where it may undergo a further local maximum
followed by a minimum while climbing the asymptotic giant branch. For masses
below about 2\,\Msun~(where the situation is very dependent on metallicity, and on how core
convective overshooting is modelled; see \S3 and Appendix A), evolution along
the first giant branch is fairly slow and proceeds to a large radius, followed
by degenerate helium ignition and a retreat in radius to the horizontal branch,
which is the low-mass analogue of the GK-giant clump stage for more massive
stars.

Most of the giants in our selection are likely to be in the GK-giant clump
because (a) that tends to be a relatively long-lived phase compared with the
first giant branch, at least provided the helium ignition phase is
non-degenerate (as is expected for masses greater than $\sim 2$\,\Msun), and (b)
GK-giant clump stars and their main-sequence companions, if they are comparable
in mass, are likely to be much more nearly equal in luminosity (and therefore
more easily recognizable as composite-spectrum binaries) compared to systems
comprising more luminous stars on the first giant branch and main-sequence
companions.  Over a substantial range of mass (2--5\,\Msun) the long-lived
minimum radius in the GK-giant clump is about 10--30\,\Rsun, and many giants in
our sample have radii in that range.

Because our modelling includes tidal friction, and mass loss through stellar wind,
we have to start the evolution of a binary with different masses,
orbital period and eccentricity from those that currently pertain. We also have
to start with a zero-age rotation period, and usually adopt 2\,d for each
component.  This paper does not make a serious attempt to solve the set of
equations that might yield more precise starting values, for three reasons:
(a) most of the current masses are not usually known to the 3\% precision of
the \citet{Torres.Andersen.2010} sample, (b) the extreme non-linearity of the
problem would probably introduce many spurious difficulties, and (c) it was in
most cases not difficult to guess a set of starting values that would be
adequate, though one might seek to improve them by iteration. There are also
several qualitative constraints: (i) the absence (or presence) of substantial
eccentricity is often a strong hint as to whether the star has (or has not)
been through its local maximum radius at helium ignition, (ii) circularisation
by tidal friction is only likely to become important if the radius of the star
exceeds about a third of its Roche-lobe radius, as seen in double-main-sequence
binaries (Pols et al. 1997), and (iii) if a giant has a circular orbit, but its
radius is less than (say) a quarter of its Roche-lobe radius, then that might
be an indication that the radius has been substantially greater in the past,
and therefore that the star {\it has} passed through helium ignition.

\def\sc{\scriptstyle}
\def\th{\thinspace}
\def\tls{{\rlap{\raise 0.5ex\hbox{$\sc  {<}$}}{\lower 0.3ex\hbox{$\sc  {\sim}$}} }}
In the case of $\zeta$~Aur (Fig.~\ref{fig:fig1}), the observational
uncertainties in radius, temperature and luminosity are too large to exclude
definitely four out of five possible solutions. The primary in the model is
almost exactly at the observed radius for the temporary maximum at helium
ignition. It will be very near the observed radius just before and just
after helium ignition; it then returns to that same radius on the asymptotic
giant branch after a truncated `blue loop', and it will in fact pass through
the same radius three times as it climbs the asymptotic giant branch. It
might have been possible to break the degeneracy by appealing to the
circularity (or otherwise) of the orbit.  The eccentricity of $\sim$0.4 of
$\zeta$~Aur's orbit might suggest that there has not yet been much tidal
interaction, but panel (c) shows that if the system commenced with $e=0.85$,
tidal friction would wear it down to $\sim$0.4 during helium ignition, after
which it would remain fairly constant for a substantial time until the primary
returned to about the same radius as in its earlier local maximum.

The model of $\alpha$~Aur (Capella; Fig.~\ref{fig:fig1}) seems to fit the
observations very well, but there are inconsistencies in the {\it latter}.  Two
recent published measurements of $K_2$ (the RV amplitude of the secondary)
appear quite precise according to their respective internal standard deviations,
but the values differ from one other by many $\sigma$: $K_2 = 26.27\pm 0.09$
\citep{Torres.Claret.2009}, or $26.840 \pm 0.024$ \citep{Weber.Strassm.2011},
equivalent to differences of 6\,$\sigma$ or 24\,$\sigma$, respectively.  In fact
our models for Capella fit much better the values of Weber \& Strassmeier.
Recently Torres et al.~(2015) have revised their $K_2$ to $26.86\pm 0.02$, in
good agreement with Weber \& Strassmeier (and our theoretical model).

For both binaries, the models include a certain amount of mass loss by way of
stellar wind, as indicated by the middle panels of Fig.~\ref{fig:fig1}. Three
types of mass loss are modelled: \\ (1) In the very reasonable expectation that
{\it all} stars, whether single or in a widish binary, with a mass less than
$\sim 8\,$\Msun\  end up as white dwarfs, we impose
a rate (referred to as `Single Red-Giant Wind')
which is assumed to be (a) proportional to the ratio of the luminosity to the
binding energy of the envelope, and (b) of sufficient strength to reduce a
non-rotating single 4\,\Msun\ star to a white dwarf of $\sim$1\,\Msun, \\ (2) a
Dynamo-Driven Wind \citep{Eggleton.2001,Eggleton.2006}, which is included
through a formulation that gives, {\it inter alia}, the mass-loss rate as a
function of rotation rate, luminosity, radius and mass (see \S~3.2), and \\ (3) a
mass-loss rate that has been determined empirically by \citet{deJager.al.1988}
for luminous stars ( $\log L \geq 4.60$), though it only applies to one or two
of our sample.

For $\zeta$ Aur the modelled mass loss is mainly by single red-giant wind,
while for $\alpha$~Aur it is mainly by dynamo-driven wind, though in neither
case is the rate high enough to affect very strongly the agreement with
observation.  The agreement is actually somewhat better with dynamo-driven wind
than without it, but it is difficult to establish that in the face of the
uncertainties in the observational data.  It may also be worth mentioning that
the chromospheric material of $\zeta$~Aur (as isolated near to occultation of
the hot star at eclipse phases) is rather tightly confined, somewhat suggestive
of a magnetic-loop formation (Dr R. E. M. Griffin, p.c.).  Furthermore, a few
of the systems in the sample show a marked mass anomaly in the sense that the
primary is {\it less} massive than the secondary, and that could {\it only}
realistically result from some level of dynamo-driven wind.

\begin{figure*} 
\centering
\begin{tabular}[]{lll}
\hskip-10pt
\includegraphics[height=6.0cm]{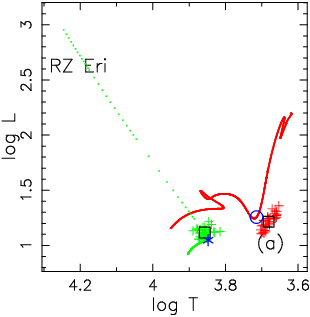} &
\hskip-10pt
\includegraphics[height=6.0cm]{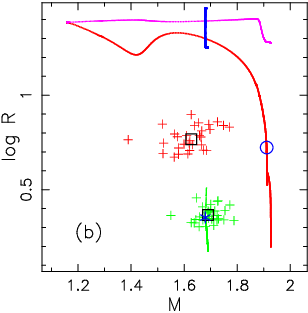} &
\hskip-10pt
\includegraphics[height=6.0cm]{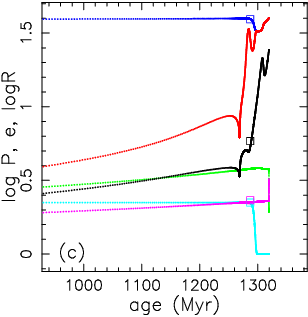} \\
\hskip-10pt
\includegraphics[height=6.0cm]{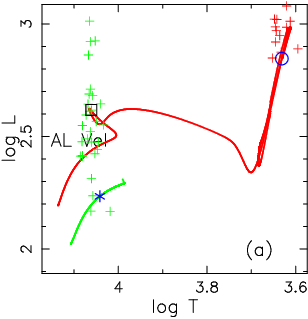} &
\hskip-10pt
\includegraphics[height=6.0cm]{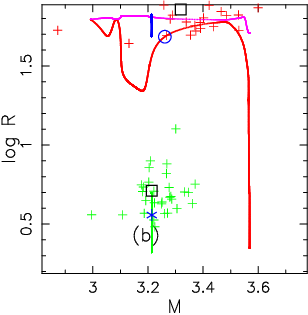} &
\hskip-10pt
\includegraphics[height=6.0cm]{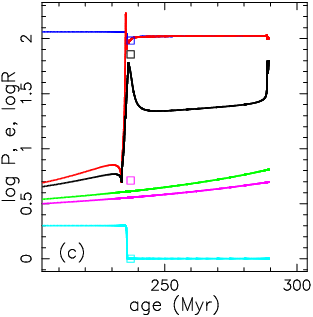} \\
\hskip-10pt
\includegraphics[height=6.0cm]{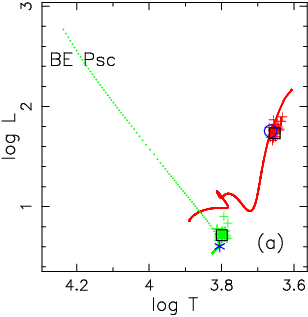} &
\hskip-10pt
\includegraphics[height=6.0cm]{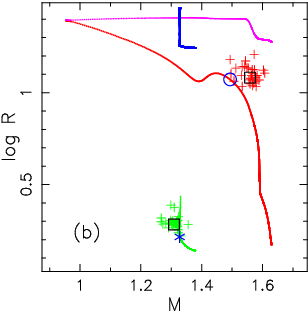} &
\hskip-10pt
\includegraphics[height=6.0cm]{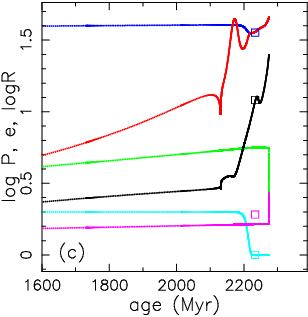} \\
\end{tabular}
\caption{Evolutionary models for RZ~Eri (top), AL~Vel (middle) and BE Psc
(bottom).  The panels and the symbols correspond to those in Fig.~\ref{fig:fig1}.
The primary of RZ~Eri is less massive than its secondary, so it must have lost
substantial mass, probably by dynamo-driven wind; however, the algorithm used
here underestimates by a factor of 30 or so the amount of mass that is lost.
The model for AL~Vel shows considerable scatter, mainly in the parallax and
hence the luminosities, but the theoretical luminosities are within the scatter.
BE~Psc seems to have undergone {\it less} mass loss than the model suggests; yet
its  parameters are rather similar to RZ~Eri.}
\label{fig:fig2}
\end{figure*}

Of the three systems illustrated in Fig.~\ref{fig:fig2}, it seems very likely
that some kind of mass loss has played a role in RZ~Eri, though it is less clear
for AL~Vel and BE~Psc.  Several other systems, such as RU~Cnc and AS-010538,
reveal either substantially more or substantially less mass loss than the
Dynamo-Driven Wind model predicts. We discuss these systems more fully in \S 5.2.

There are several red-giant+main-sequence binaries which are semi-detached.
They have not been included in the sample, as most are of fairly short period
and we have set a limit at $P\leq$8 d.  Longer-period ones such as SS~Lep
\citep[$P$ = 250\,d;][]{Blind.al.2011} are symbiotic binaries, and have
parameters that are potentially interesting, but they present complications which
render precise analyses difficult; they have also been excluded from our sample.

\section{Features of The Theoretical Model}

Certain elements of a stellar evolution code can be regarded as fairly
standard; they include the equation of state, the nuclear reaction network,
hydrostatic equilibrium, and the radiative opacity (though see below).
However, other elements can vary significantly from one code to another because
a soundly-based physical model is not available.  That is true for \\(i)
convection, with the mixing-length theory being normal but not necessarily
accurate, \\ (ii) semi-convective mixing -- the formulation adopted here is a
very simple diffusive approximation \citep{Eggleton.1972}, \\ (iii) convective
core overshooting, \\(iv) stellar-wind mass loss, including wind that is driven
by dynamo action owing to rapid rotation, as in RS~CVn stars, and other
mass-loss mechanisms that would reduce a single red giant to a white dwarf as
it is evolved towards the top of the asymptotic giant branch by Single Red
Giant Wind\\
(v) tidal friction, that compels giants in binaries to rotate
much more rapidly than if they were single, and which also tends to circularise
orbits that were initially eccentric,\\(vi) rotationally-driven mixing,
and \\(vii) diffusive separation of abundances.

The code used here does not incorporate elements (vi) and (vii), mainly because
it is conjectured that they will not be very important for the long-term
evolution of the stars in the sample.  There is no doubt that the surfaces of
certain A or F stars can be affected by diffusive separation, leading to Ap, Am
and Fm abundance anomalies, but the diffusion is believed to be confined to
near-surface layers and is rapidly reversed once a star crosses enough of the
Hertzsprung gap for the outer few per cent by mass to be mixed more deeply (as
in the case of o~Leo, \S4.2).  Where diffusive separation might make a
difference in the long term is in stars of about 1\,\Msun, where nuclear
evolution is sufficiently slow that diffusion might separate helium and
hydrogen significantly in the deep interior. However, very few of the components
considered here have masses $<$ 1.5\,\Msun. Rotationally-driven mixing has been
proposed for early type stars, but \citet{Tkachenko.al.2014} found no
evidence for it in a detailed abundance analysis of V380~Cyg (\S4.2).

The code used here adopts the opacities of \citet{Rogers.Igles.1992}.
\citet{Asplund.al.2000,Asplund.al.2005} have suggested that, on the basis of
3-D modelling of the Sun's convective zone and photosphere, the solar
metallicity is somewhat less than the previously standard value of $Z=0.02$,
but -- as maintained by \citet{Basu.Antia.2006} -- it has so far proved hard to
reconcile that claim with the previous good agreement between
helioseismological results based on the `standard' metallicity
\citep[e.g.][]{C-Dalsgaard.Dappen.1992}.  In the meantime we are continuing to
use the standard metallicity.

\par We use an implicitly adaptive mesh-point distribution (Eggleton 1971) which
allows us to model stars with no more than 200 meshpoints in them, from centre to
photosphere, even with double shell burning. This economy is counterbalanced by
the fact that we choose to solve 44 difference equations simultaneously. For
example, we solve Clairault's equation (a second-order DE) for the distortion
of each component along with two other first-order DEs that determine the tidal
velocity field and the rate of its dissipation by turbulent convective viscosity.
The code runs easily on an Apple Mac Pro (reconfigured for Linux, with a Fortran
compiler), and takes between 10 minutes and about an hour to solve each of the
60 systems.

The following subsections discuss, in turn, convective core overshooting, wind
mass loss, and tidal friction.

\subsection{Core Convective Overshooting}

The model for core convective overshooting, based here on that proposed by
\citet{Eggleton.2006}, assumes that mixing in the core goes beyond the
Schwarzschild boundary ($\nabla_r-\nabla_a=0$) to a boundary
$\nabla_r-\nabla_a=-\dOS < 0$.  The functional form of $\dOS$ may ultimately be
determined by 3-D numerical simulations, but more than $10^{12}$ mesh-points
will be necessary and such refinement has probably not yet been reached.  It is
to be hoped that the 1-D modelling presented here places some restrictions
on $\dOS$. In particular, the models for TZ~For, SU~Cyg, V380~Cyg and $\delta$~Ori,
which have primary masses of $\sim$2, 6, 13 and 24\,\Msun, respectively, show that
a modest amount of overshooting must operate between 2 and 6\,\Msun~but that by
13\,\Msun~the amount (measured in pressure scale-heights, PSH) must be trebled,
and even quadrupled by 24\,\Msun.
The functional form used is given in Appendix A; its effect is to create mixing
over an extra 0.16--0.2 PSH in stars with masses $\lesssim $\,4\,\Msun, and over
0.5--0.7 PSH for masses of about 10--13\,\Msun; the region affected may in fact
extend to $\sim$1 PSH by 40\,\Msun, but that condition has not yet been tested.
It should be noted that the model described and used here differs a little from
those used in earlier versions of the same code (e.g.,~by Pols et al.~1997) by
including modestly more core convective overshooting for lower masses, and
substantially more for higher masses (as in V380~Cyg and $\delta$ Ori).
\begin{figure*}  
\centering
\begin{tabular}[]{lll}
\hskip-10pt
\includegraphics[height=6.0cm]{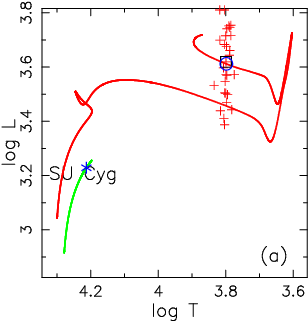} &
\hskip-10pt
\includegraphics[height=6.0cm]{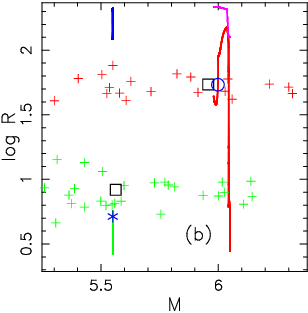} &
\hskip-10pt
\includegraphics[height=6.0cm]{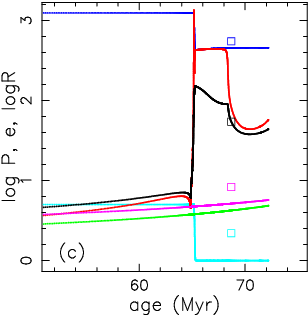} \\
\includegraphics[height=6.0cm]{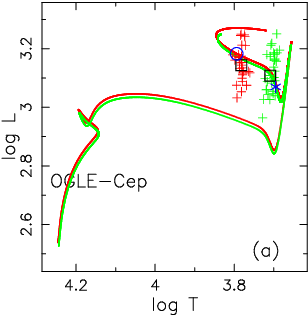} &
\hskip-10pt
\includegraphics[height=6.0cm]{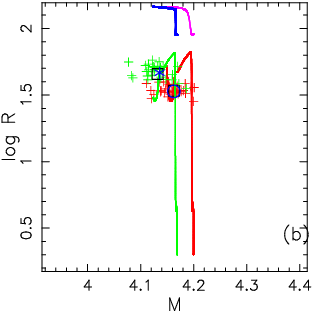} &
\hskip-10pt
\includegraphics[height=6.0cm]{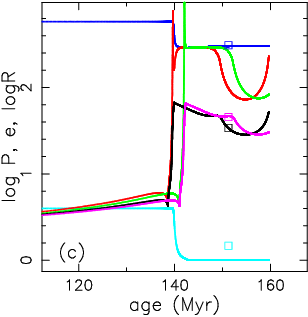} \\
\includegraphics[height=6.0cm]{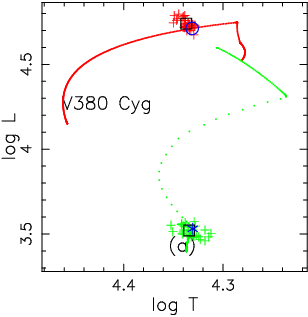} &
\hskip-10pt
\includegraphics[height=6.0cm]{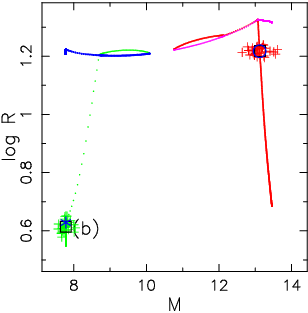} &
\hskip-10pt
\includegraphics[height=6.0cm]{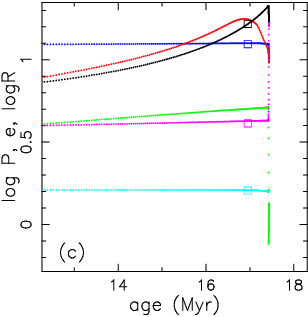} \\
\end{tabular}
\caption{Upper panels: the Cepheid SU~Cyg, treated as a binary (although it is
actually a triple).  The `secondary' is a fictitious object with the same mass
as the combined sub-binary mass.  Middle panels: the Cepheid binary OGLE-Cep in
the LMC; $*2$ is in the GK giant clump. Lower panels: V380~Cyg, where $*1$ is a
very large star that would be well beyond the upper edge of the main-sequence
band unless it had very considerable core convective overshooting.}
\label{fig:fig3}
\end{figure*}

TZ~For is critical to this discussion because it seems clear that the primary
star ($*1$) must have passed through non-degenerate helium ignition. That would
explain its circular 76-d orbit despite the fact that $*1$ is less than 20\%
of its Roche-lobe radius. Without overshooting, for masses below 2.5\,\Msun~the
helium ignition would be a degenerate He flash, requiring $*1$ to reach a much
bigger radius and hence undergo substantial Roche-lobe overflow.  If the red
giant in TZ~For were on the first giant branch, it would not yet be large
enough to circularize the orbit; however, if it is in the GK-giant clump it
must have undergone non-degenerate helium ignition at a modest radius that was
two or three times larger than its present one ($\sim$8.5\,\Rsun) but smaller
than its Roche-lobe one ($\sim$45\,\Rsun).  DQ~Leo, $\alpha$~Equ and $\eta$~And
reveal similar evidence, having only slightly different masses and period, and
circular orbits.

Primaries in the GK-giant clump that are more massive than about 2.5\,\Msun~are
not quite so informative, because they would undergo non-degenerate helium
ignition either with or without overshooting. They may nevertheless present
more information about tidal friction (\S3.2).  A star in the GK-giant clump
with a mass of about 6\,\Msun~starts to evolve towards the blue and into the
blue loop, where it may be conspicuous as a Cepheid.  Reconciling theoretical
Cepheid blue loops with observation was a problem for a long time, but was
largely resolved by incorporating overshooting into the models
\citep{Schroder.al.1997}.

\setcounter{table}{1}
\begin{table*}
\caption{Hypothetical Mass Loss and Dynamo Activity during the Sun's Evolution
\label{tab:2new}}
 \indent  \\
\begin{tabular}[]{rllllrllll}
   n  &age      & $M$     &$P_{rot}$&$\log R$&$\log L$&\ \ \ \  $\dot{M}$  &$B_P$ &$R_A/R$&  \\
      & Gyr     &\Msun    &  d    &\Rsun  &\Lsun   &  \Msun/yr             &Gauss &       &\\
    3 & 0.000   & 1.0242  & 36.71 &  1.019&  1.519 &  $2.5\times 10^{-8} $ & 15.1 &  1.70 & Arbitrary starting point on the Hayashi track\\
 1004 & 0.042   & 1.0129  & 2.991 & -0.050& -0.137 &  $5.3\times 10^{-11}$ & 20.8 &  2.65 & Minimum radius, at ZAMS\\
 1110 & 0.278   & 1.0044  & 6.731 & -0.045& -0.125 &  $2.0\times 10^{-11}$ & 14.5 &  3.21 & Rotation slowed, mass loss much down\\
 1202 & 4.567   & 0.9999  & 24.89 &  0.000&  0.005 &  $3.1\times 10^{-14}$ & 1.29 &  11.3 & Present day\\
 1400 & 10.30   & 0.9970  & 47.38 &  0.166&  0.307 &  $1.1\times 10^{-12}$ & 0.35 &  0.35 & Hertzsprung gap\\
 2200 & 11.77   & 0.9861  & 2949  &  0.887&  1.384 &  $1.1\times 10^{-10}$ & 0.0  &  0.0  & Lower first giant branch\\
 2360 & 11.87   & 0.9379  &   -   &  1.523&  2.362 &  $3.4\times 10^{-9} $ &  -   &   -   & Single red-giant wind becoming significant\\
 2566 & 11.88   & 0.6224  &   -   &  2.367&  3.405 &  $1.1\times 10^{-7} $ &  -   &   -   & He flash\\
 2567 & 11.88   & 0.6207  &   -   &  1.027&  1.705 &  $8.4\times 10^{-11}$ &  -   &   -   & `Zero Age' Horizontal Branch\\
 2976 & 11.97   & 0.6146  &   -   &  0.924&  1.634 &  $5.9\times 10^{-11}$ &  -   &   -   & Local minimum radius\\
 3809 & 12.04   & 0.5525  &   -   &  1.934&  3.388 &  $3.6\times 10^{-10}$ &  -   &   -   & Tip of AGB.\\
\end{tabular}
\end{table*}
Masses for Cepheids have rarely been obtained directly from double-lined
eclipsing (or interferometric) orbits. However one such system in the LMC,
OGLE-Cep (see Table 3), has been found to have
parameters of 4.165 + 4.134\,\Msun, 309.4\,d, $e$ = 0.166
\citep{Pilecki.al.2013}.  The system can be fitted very easily by a theoretical model
(Fig.~\ref{fig:fig3}), but it needs to use a metallicity that is substantially
less than solar. An increase in metallicity tends to reduce the size of blue
loops rather drastically. At solar metallicity, blue loops large enough to
produce Cepheids are confined to masses greater than $\sim$5.5\,\Msun, but it
also depends on the degree of assumed overshooting; too much shrinks the blue
loop to insignificance.  We estimate that overshooting at $\sim$6\,\Msun,
roughly the mass of the double-lined but non-eclipsing Cepheid SU~Cyg
\citep{Evans.Bolton.1990}, must be not much more than at $\sim$2\,\Msun~(as in
TZ~For).

It is interesting to note that the companion to SU~Cyg is itself a fairly
compact sub-binary of period 4.65\,d.  Fig.~3 models the SU~Cyg system with a
fictitious secondary component ($*2$) that has the same mass as the sub-binary.
The primary develops a blue loop that gets it to the location of the
Cepheid, though at higher masses still (as in V380~Cyg) it is necessary to
include substantially greater overshooting.  However, both those Cepheids
present a problem, inasmuch as both have eccentric orbits and yet both should
have circularized their orbits (according to our models) during the helium
ignition stage when the components were larger by a factor of two or more. This
is discussed further in \S5.3.

V380~Cyg is not an obvious candidate for the present study, since although $*1$,
at B1.5\,III, is technically a giant, it is very much bluer than almost all the
other giants. We would argue that $*1$ {\it must} be still within the main
sequence band, because if it were in the Hertzsprung gap it would be evolving
very rapidly, on a timescale of $\sim$100\,yr. By contrast, if it is still
in the main sequence band (Fig.~3) its evolutionary timescale is more like
$10^4\,$yr. This system's relevance to overshooting has been discussed by
several authors, including \citet{Pietrzynski.al.2009}. Also $\delta$~Ori is
an atypical addition: it has an O9.5\,II primary, which nevertheless must
(we think) be still in the MS band for the same reason.

\subsection{Dynamo-Driven Wind  and Single Red Giant Wind}

As mentioned above, stellar wind mass loss may be regarded as a combination of
three contributions. Probably the most significant one for the stars in our
sample is dynamo-driven wind, a model for which is discussed in some detail by
\citet{Eggleton.2001,Eggleton.2006}. From an input of mass, radius, luminosity
and stellar rotation period this model produces estimates for (a) the
differential rotation rate between the convective envelope and the radiative
core, (b) the star-spot cycle time (e.g., 22 years for the Sun), (c) the
overall poloidal magnetic field, (d) the mass-loss rate (assuming that the mass
loss is driven by destruction of the toroidal field at and above the surface of the
star), and (e) the Alfv\'en radius of the wind as determined by the poloidal
field and the wind strength.  The rotation rate will modify itself in the
course of time through magnetic braking, whereby angular momentum is
transferred to the wind; the latter is assumed to be rotating rigidly out to
the Alfv\'en radius and then escaping freely.  This process works for single
stars as well as stars in binaries, though in single stars it is self-limiting
because the dynamo weakens as the star spins down, whereas in binary stars that
are close enough it can be self-amplifying, since tidal friction may reduce the
separation and therefore the spin rate increases as the star loses angular
momentum to the wind.

Table~2 gives a few stages in the evolution of a single star that
resembles the Sun at 4.567~Gyr. It tabulates the rotational period,
the mass-loss rate, the poloidal magnetic field and the Alfv{\'e}n radius. The
Table suggests that a dynamo-driven wind is only responsible for
significant mass loss in roughly the first 300\,Myr; most occurs in just the first
150\,Myr, by which time the rotation has slowed to about 5\,d from a peak value
of 3\,d.  Subsequent mass loss, producing a white-dwarf precursor of
0.55\,\Msun, is modelled by a `Reimers-like' wind \citep{Reimers.1975}, where
$\dot{M}$ is proportional to the ratio of luminosity to the binding energy of
the envelope above the burning shell, as described in \S2 above and referred to
as a `single red-giant wind'.

Mass loss through a dynamo-driven wind affects all of our theoretical models in
principle, but in the great majority it makes rather little difference.  The
three systems represented in Fig.~2 display a range of disparity in the inferred
rates of mass loss ranging from about $\sim$20 times more than is predicted
for RZ~Eri to $\sim$3 times less than predicted for BE~Psc. For AL~Vel
the predicted amount of mass loss appears to match what can be inferred from
observation to within a factor of $\sim$2. In HR~6046 (online only) the
theoretical mass loss exceeds what is probably required by a factor of about 10.

   Several (11) of our 60 systems come into substantial conflict with our mass-loss
algorithms. We discuss these individually in \S 4.2 and collectively in \S5.2.

\subsection{Tidal Friction.}

The model of tidal friction used here has been described in some detail by
\citet{Eggleton.2006}, and in a somewhat preliminary version by
\citet{Eggleton.al.1999}. It relies on turbulent convective motion as the
dissipatory agent for tidal motion. For the most part it seems to be effective
at circularizing orbits that are known to be circular now, but which are wide
enough that they were very probably eccentric at age zero -- as in the case of
$\alpha$~Aur (Fig.~1). In that system the primary is about 8 times smaller than
its Roche lobe, and tidal friction is unlikely to have circularized its orbit unless
its radius were about 3 times its present size at helium ignition (based on a
comparison with double-main-sequence binaries).  In
$\zeta$~Aur (Fig.~1) the orbit, still eccentric ($e \sim$ 0.4), can be
modelled satisfactorily by adopting an initial $e$ = 0.85; the model suggests
that it became partly circularized during helium ignition, $e$ fell to its present
level, and will drop fairly rapidly in the future.

  Only 4 systems come into substantial conflict with our tidal-friction model.
We discuss these in \S5.3.

\begin{figure*}  
\centering
\begin{tabular}[]{lll}
\hskip-10pt
\includegraphics[height=6.0cm]{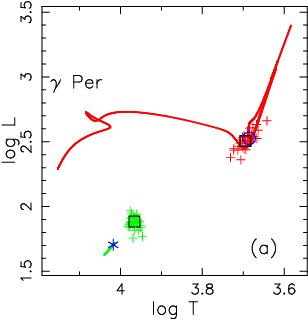} &
\hskip-10pt
\includegraphics[height=6.0cm]{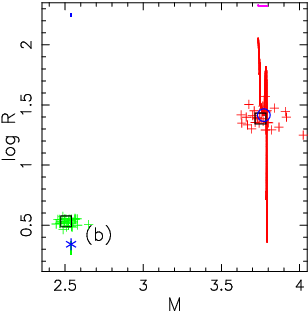} &
\hskip-10pt
\includegraphics[height=6.0cm]{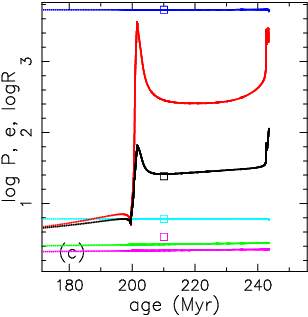} \\
\hskip-10pt
\includegraphics[height=6.0cm]{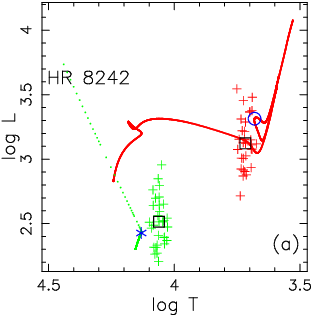} &
\hskip-10pt
\includegraphics[height=6.0cm]{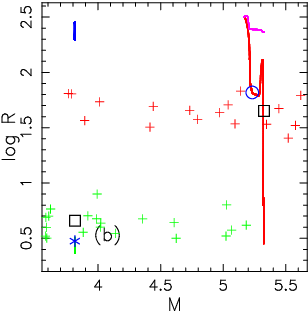} &
\hskip-10pt
\includegraphics[height=6.0cm]{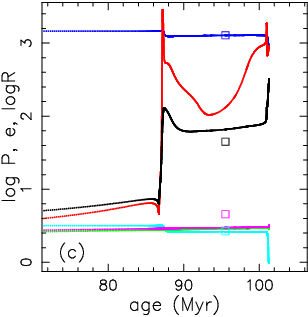} \\
\end{tabular}
\caption{Evolutionary models for $\gamma$~Per (top) and HR~8242 (bottom). The
panels and symbols correspond to those in Fig.~\ref{fig:fig1}. Both systems
have secondaries that have evolved quite a long way across the main-sequence
band, whereas the co\ae val points on their tracks are near the ZAMS. One
suggestion is that in both systems the giant is the product of a merger of what
used to be a sub-binary.}
\label{fig:fig4}
\end{figure*}
\subsection{`Over-Evolved Secondaries'}
Fig.~\ref{fig:fig4} shows our attempts to model $\gamma$~Per (upper set) and
HR~8242 (lower set). In both systems the secondary appears to have evolved {\it
considerably} more than it could have done in the time that the primary took to
reach something like its present radius. The observed mass ratio is about 1.5
in both cases and the secondary should have barely left the ZAMS, but in fact
it has evolved to something like twice its ZAMS radius.  One explanation could
be that the binary was formed by a capture process between an older star and a
younger star, but it seems very unlikely that this occurred in two out of 60
systems.  A different, and possibly more tenable, explanation is offered in
\S5.1.
\section{Individual Cases}\label{sect:cases}

\subsection{Presenting the information}

The sample of 60 systems was listed in Table~1, together with some aliases and
the primary literature references.  Table~3 records what has been found in the
literature  about each system from radial velocity measurements  of both
components, from modelling the photometry and spectroscopy, and from astrometry.
For each system ten or eleven more-or-less directly measured quantities,
which we refer to as `raw', are
listed on the first line, with their measurement uncertainties on the second line.
The quantities range from orbital radial velocity amplitudes to parallax, for each
system. The eleventh measurable quantity, inclination, is of course not
available unless the system is either eclipsing or astrometric. These quantities
are transformed by a standard procedure (Appendix B) into quantities which we
refer to as `derived' --
mass, temperature, radius and luminosity, for each component -- that are easily
compared with the theoretical models, and that are are listed in Table 4.
Each system is illustrated by a plot consisting of three panels, as for the 10
systems in Figs~1~--~4. The 50 not shown here are accessible online.
\def\pn{\par\noindent}

Since a spectral type is a visual description of a spectrum rather than a
measurement of it, and since the isolated spectrum of the primary cannot be
seen in most of these binaries, there is unavoidably some degree of
subjectivity attached to the spectral types listed in Table~3.
The spectral type of the primary is usually deemed to be that of the standard
which was adopted as its surrogate in the subtraction procedure to uncover
the secondary spectrum, though not
infrequently (particularly for the brighter giants) the match can be
less than ideal.  For the secondary, the individualities of available single,
and preferably low-rotating, standard early-type spectra present a different
challenge and may be circumnavigated by fitting a synthetic spectrum to the
extracted (supposedly pure) version of its spectrum which then has to be
translated into a spectral type, often (also somewhat subjectively) via its
($B-V$) as tabulated by (for instance) \citet{Kaler.1982}.  The spectral types
listed in Table~3 are therefore guides rather than accurate statements.

The tabulated parallaxes are mostly either the re-worked Hipparcos values
\citep{Leeuwen.2007} or else from Gaia \citep{gaia1,gaia2}; but in principle a system that is both eclipsing and
double-lined can provide a parallax independent of astrometry, as for several
LMC and SMC systems.  It should be noted that a system with an orbital period
of order one year may have an inherently ambiguous astrometric parallax as a
result of the confusion of the target's orbit and its parallactic motion.

The data compiled in Table~3 consists of 10 or 11 observationally determined
numbers per system. Many
systems have an observed inclination, as determined by either an eclipse or an
astrometric orbit or both, but several do not, and we then estimate an
inclination by matching the system to theoretical systems. In Table~3 an E (39),
A (13) or N (10) in the last column  means Eclipsing, Astrometric or Neither.
Two are both E and A.
\setcounter{table}{2}
\begin{table*}
\tiny
\setlength{\tabcolsep}{0.075in}
\caption{Observed and modified quantities
\label{tab:3new}}
\indent
\begin{tabular}[]{rlrcrrrrrrrrrcr}
\hline
No.     & Name                  &  $P$          & $e$           & $K_1$         & $K_2$         & \V12          &   \DV         & \AV           &\Ta~\,         &\Tb~\,         & plx            &$i$   & Type  \\
        &  spectra              &  day          &               &  km/s         &  km/s         &               &               &               &               &               & mas            &      & GoF   \\
        &                       &               &               &               &               &               &               &               &               &               &                &      & Z     \\
\hline
 1&	SMC-130	        &       120.470	&	0.000	&	33.42	&	32.54	&	16.783	&	-0.72	&0.24	&4515	&	4912	&       0.0162   &83.09	&E	\\
  &	G7III   	&	0.001	&	0.000	&	0.12	&	0.11	&	0.010	&	0.10	&0.02	&150	&	150	&	0.0008   &0.10	&1.08	\\
  &	 + K1III	&	120.470	&	0.000	&	33.42	&	32.54	&	16.783	&	-0.95	&0.24	&4365	&	4812	&	0.0180   &83.09	&.004	\\
 2&	SMC-126   	&	635.000	&	0.042	&	18.48	&	18.54	&	16.771	&	-0.192	&0.24	&4480	&	4510	&	0.0160   &86.92	&E	\\
  &	K2III     	&	0.009	&	0.002	&	0.110	&	0.10	&	0.01	&	0.020	&0.02	&150	&	150	&	0.0030   &0.09	&0.92	\\
  &	 + K1III        &	635.000	&	0.042	&	18.48	&	18.54	&	16.771	&	-0.222	&0.24	&4250	&	4350	&	0.0165   &86.92	&.004	\\
 3&	SMC-101 	&	102.900	&	0.000	&	39.44	&	41.03	&	17.177	&	-0.203	&0.20	&5170	&	5580	&	0.0154   &88.04	&E	\\
  &	K2III           &	0.000	&	0.000	&	0.20	&	0.12	&	0.010	&	0.020	&0.02	&95	&	90	&	0.0003   &0.23	&1.05	\\
  &	 + K1II         &	102.900	&	0.000	&	39.44	&	41.03	&	17.177	&	-0.203	&0.20	&5170	&	5280	&	0.0154   &88.04	&.004	\\
 4&	HD 4615	        &	302.771	&	0.435	&	27.52	&	30.8	&	6.82	&	-1.10	&0.25	&4400	&	8700	&	2.48	 &71.4	&N	\\
  &	K2III           &	0.020	&	0.003	&	0.10	&	0.9	&	0.02	&	0.20	&0.05	&200	&	500	&	0.59	 &2.0	&0.17	\\
  &	 + A2V          &	302.771	&	0.435	&	27.52	&	30.8	&	6.82	&	-1.10	&0.25	&4400	&	8500	&	2.68	 &71.4	&.02	\\
 5&	$\eta$ And	&	115.73	&	0.003	&	17.98	&	19.03	&	4.40	&	-0.54	&0.05	&5050	&	5000	&	13.3	 &30.5	&A	\\
  &	G8III           &	0.02	&	0.002	&	0.09	&	0.11	&	0.02	&	0.02	&0.01	&200	&	200	&	0.5	 &0.20	&0.40	\\
  &	 + G8III	&	115.73	&	0.003	&	18.04	&	18.92	&	4.40	&	-0.54	&0.05	&5000	&	5050	&	13.3	 &30.5	&.02	\\
 6&	SMC-108 	&	185.220	&	0.000	&	37.85	&	37.96	&	15.205	&	0.081	&0.28	&4955	&	5675	&	0.01562  &78.87	&E	\\
  &	F9II + G7II	&	0.002	&	0.000	&	0.08	&	0.09	&	0.01	&	0.02	&0.02	&105	&	90	&	0.00030  &0.10	&0.00	\\
  &	 + G7III        &	185.220	&	0.000	&	37.85	&	37.96	&	15.205	&	0.081	&0.28	&4955	&	5675	&	0.01562  &78.87	&.004	\\
 7&	BE Psc	        &	35.670	&	0.000	&	41.52	&	49.43	&	8.76	&	-1.95	&0.80	&4500	&	6300	&	3.8	 &81.8	&E	\\
  &	K1III           &	0.000	&	0.000	&	0.19	&	0.41	&	0.02	&	0.05	&0.02	&70	&	100	&	0.2	 &0.1	&1.05	\\
  &	 + F6IV-V       &	35.670	&	0.000	&	41.52	&	49.24	&	8.76	&	-2.25	&0.80	&4550	&	6350	&	3.8	 &81.8	&.02	\\
 8&	ASAS--010538    &	8.069	&	0.000	&	73.0	&	75.74	&	10.1	&	-0.75	&0.19	&4889	&	6156	&	2.66	 &79.90	&E	\\
  &		        &	0.000	&	0.000	&	1.3	&	0.26	&	0.2     &	0.05	&0.02	&98      &	176	&	0.23	 &0.65	&0.45	\\
  &		        &	8.069	&	0.000	&	73.0	&	75.74	&	10.1	&	-0.75	&0.19	&4889	&	6156	&	2.35	 &79.90	&.02	\\
 9&	AI Phe	        &	24.590	&	0.188	&	49.24	&	50.90	&	8.58	&	0.22	&0.01	&5010	&	6310	&	5.94	 &88.45	&E	\\
  &	K0IV    	&	0.000	&	0.000	&	0.10	&	0.10	&	0.02	&	0.25	&0.05	&250	&	300	&	0.24	 &1.00	&0.64	\\
  &	 + F7V	        &	24.590	&	0.188	&	49.24	&	50.90	&	8.58	&	-0.22	&0.01	&4910	&	6110	&	5.78	 &88.45	&.01	\\
10&	$\tau$ Per	&	1515.880&	0.734	&	19.09	&	23.0	&	3.93	&	-1.71	&0.00	&5050	&	8000	&	12.83	 &85	&EA	\\
  &	G8IIIa          &	0.100	&	0.007	&	0.35	&	4.0	&	0.02	&	0.10	&0.05	&150	&	300	&	.36	 &2	&0.89	\\
  &	 + A6V	        &	1515.880&	0.734	&	19.09	&	23.3	&	3.93	&	-1.81	&0.00	&4950	&	8100	&	13.70	 &85	&.01	\\
11&	$\gamma$ Per    &	5327.7  &	0.785	&	14.53	&	21.73	&	2.91	&	-1.33	&0.00	&4950	&	9250	&	13.41	 &90	&EA	\\
  &	G8IIIa          &	0.6	&	0.002	&	0.10	&	0.20	&	0.02	&	0.10	&0.05	&200	&	200	&	0.51	 &2.00	& --	\\
  &	 + A2IV	        &	5327.7  &	0.785	&	14.53	&	21.73	&	2.91	&	-1.33	&0.00	&4950	&	9250	&	13.41	 &90.00	&.02	\\
12&	TZ For	        &	75.770	&	0.000	&	38.81	&	40.80	&	6.89	&	-0.25	&0.17	&4950	&	6350	&	5.44	 &85.6	&E	\\
  &	G8III           &	0.000	&	0.010	&	0.01	&	0.02	&	0.02	&        0.10	&0.05	&200	&	250	&	0.25	 &0.05	&0.57	\\
  &	 + F7III        &	75.666	&	0.000	&	38.81	&	40.80	&	6.89	&	-0.25	&0.17	&4900	&	6550	&	5.75	 &85.6	&.03	\\
13&	HR 1129	        &	6124.   &	0.678	&	15.87	&	17.60	&	4.82	&	-1.20	&0.93	&5250	&	13000	&	3.29	 &87	&A	\\
  &	G2Ib-II         &       3.      &	0.003	&	0.08	&	0.34	&	0.02	&	0.40	&0.35	&150	&	500	&	0.27	 &4	&0.80	\\
  &	 + B7III-IV	&	6124.   &	0.678	&	15.90	&	17.40	&	4.82	&	-1.60	&0.63	&5250	&	14000	&	3.20	 &87	&.01	\\
14&     OGLE-Cep        &       309.400 &       0.166   &       32.14   &       32.38  &       15.32   &       -0.310  &0.50   &6050   &       5120    &       0.020    &86.83 &E      \\
  &     F7Ib            &       0.100   &       0.003   &       0.07    &       0.06   &       0.02    &       0.010   &0.05   &160    &        130    &       0.001    &0.02  &0.00   \\
  &      + G4II         &       309.400 &       0.166   &       32.14   &       32.38  &       15.32   &       -0.310  &0.50   &6050   &       5120    &       0.020    &86.83 &.004   \\
15&	RZ Eri	        &	39.280	&	0.350	&	50.80	&	48.90	&	7.78	&	0.20	&0.00	&4800	&	7200	&	5.55	 &89	&E	\\
  &	G8-K0III        &       0.000   &	0.010	&	0.10	&	1.00	&	0.02	&	0.10	&0.05	&200	&	300	&	0.44	 &2	&0.72	\\
  &      + A8-F0IV      &	39.280	&	0.350	&	50.80	&	48.90	&	7.78	&	0.05	&0.00	&5000	&	7000	&	5.00	 &89	&.02	\\
16&	OGLE-01866      &	251.007	&	0.241	&	33.28	&	33.27	&	16.12	&	-0.074	&0.345	&4541	&	5327	&	0.020    &83.3 	&E	\\
  &		        &	0.004	&	0.001	&	0.05	&	0.14	&	0.09	&	0.090	&0.060	&81	&	87	&	0.001    &0.10	&0.00	\\
  &		        &	251.007	&	0.241	&	33.28	&	33.27	&	16.12	&	-0.074	&0.345	&4541	&	5050	&	0.020    &83.3  &.004	\\
17&	OGLE-03160      &	150.020	&	0.000	&	30.35	&	30.47	&	17.4	&	-1.136	&0.37	&4490	&	4954	&	0.0199   &83.36	&E      \\
  &		        &	0.001	&	0.001	&	0.11	&	0.14	&	0.09	&	0.090	&0.09	&77	&	72	&	0.0010   &0.57	&0.00	\\
  &                     &	150.020	&	0.000	&	30.35	&	30.47	&	17.4	&	-1.136	&0.37	&4490	&	4954	&	0.0199   &83.36	&.004	\\
18&	$\zeta$ Aur	&	972.150	&	0.393	&	23.26	&	27.80	&	3.69	&	-2.11	&0.25	&3960	&	15200	&	4.15	 &90	&E	\\
  &	K4Ib      	&	0.060	&	0.005	&	0.15	&	2.80	&	0.02	&	0.10	&0.05	&100	&	200	&	0.29	 &2.00	&0.49	\\
  &	 + B6V          &	972.150	&	0.393	&	23.26	&	27.60	&	3.69	&	-2.11	&0.25	&3960	&	14900	&	4.25	 &90	&.02	\\
19&	OGLE-06575      &	189.822	&	0.000	&	37.72	&	36.03	&	15.712	&	-0.152	&0.32	&4681	&	4903	&	0.020    &82.06	&E	\\
  &		        &	0.002	&	0.000	&	0.07	&	0.09	&	0.090	&	0.090	&0.09	&77	&	72	&	0.001    &0.13	&0.00	\\
  &		        &	189.822	&	0.000	&	37.72	&	36.03	&	15.712	&	-0.152	&0.32	&4681	&	4903	&	0.020    &82.06	&.004	\\
20&	OGLE-EB 	&	214.400	&	0.040	&	32.65	&	33.67	&	16.2	&	-0.44	&0.01	&5288	&	5470	&	0.020	 &88.2	&E	\\
  &	K4III           &	0.001	&	0.010	&	0.08	&	0.10	&	0.02	&	0.10	&0.05	&81	&	96	&	0.001    &0.1	&0.00	\\
  &	 + K4III        &	214.171	&	0.039	&	32.76	&	33.37	&	16.2	&	-0.44	&0.01	&5288	&	5470	&	0.020	 &88.2	&.004	\\
21&	OGLE-09660      &	167.635	&	0.052	&	35.13	&	34.91	&	16.27	&	-0.504	&0.38	&5352	&	4677	&	0.020    &87.8	&E	\\
  &		        &	0.001	&	0.001	&	0.08	&	0.16	&	0.090	&	0.090	&0.01	&99	&	99	&	0.002    &0.3	&0.81	\\
  &		        &	167.635	&	0.052	&	35.13	&	34.91	&	16.27	&	-0.504	&0.38	&5152	&	4677	&	0.023    &87.8	&.004	\\
22&	OGLE-10567      &	117.871	&	0.000	&	39.31	&	41.32	&	16.48	&	-0.26	&0.30	&5067	&	4704	&	0.020    &83.4	&E	\\
  &	                &	0.001	&	0.000	&	0.14	&	0.13	&	0.09	&	0.09	&0.09	&80	&	73	&	0.001    &0.3	&0.00	\\
  &		        &	117.871	&	0.000	&	39.31	&	41.32	&	16.48	&	-0.26	&0.30	&5067	&	4704	&	0.020    &83.4	&.004	\\
23&	OGLE-26122      &	771.781	&	0.419	&	23.8	&	25.08	&	16.63	&	-0.76	&0.42	&4989	&	4995	&	0.020    &88.45	&E	\\
  &		        &	0.005	&	0.002	&	0.1	&	0.14	&	0.09	&	0.01	&0.01	&99.9	&	99.9	&	0.001    &0.04	&0.00	\\
  &		        &	771.781	&	0.419	&	23.8	&	25.08	&	16.63	&	-0.76	&0.42	&4989	&	4995	&	0.020    &88.45	&.004	\\
24&	$\alpha$ Aur    &	104.000	&	0.001	&	25.96	&	26.840	&	0.08	&	0.13	&0.01	&4920	&	5680	&	76.19	 &137.2	&A	\\
  &	G9III           &	0.000	&	0.001	&	0.00	&	0.024	&	0.01	&	0.10	&0.01	&196.8	&	230	&	0.47	 &0.05	&0.54	\\
  &	 + G0III        &	104.000	&	0.000	&	25.96	&	26.840	&	0.08	&	0.23	&0.01	&4920	&	5900	&	76.19	 & 137.2&.02	\\
25&	OGLE-15260      &	157.324	&	0.000	&	27.93	&	27.67	&	17	&	-0.676	&0.30	&4320	&	4706	&	0.0199   & 82.9 &E	\\
  &		        &	0.001	&	0.000	&	0.14	&	0.11	&	0.09	&	0.090	&0.09	&81	&	87	&	0.0010   & 0.3	&0.00	\\
  &		        &	157.324	&	0.000	&	27.93	&	27.67	&	17	&	-0.676	&0.30	&4320	&	4706	&	0.0199   & 82.9 &.004	\\
26&	$\delta$ Ori    &	5.732	&	0.112	&	104.6	&	266.0	&	2.41	&	-3.00	&0.15	&30000	&	24000	&	2.9	 & 76.4	&E	\\
  &	O9.5II          &	0.001	&	0.010	&	1.6	&	20.0	&	0.02	&	0.40	&0.05	&1000	&	1000	&	0.5	 & 0.2	&0.25	\\
  &	 + B0V 	        &	5.732	&	0.112	&	104.6	&	266.0	&	2.41	&	-3.20	&0.15	&30500	&	23700	&	2.8	 & 76.4	&.02	\\
27&	HR 2030	        &	66.452	&	0.017	&	25.98	&	25.6	&	5.96	&	-0.20	&0.40	&4550	&	11750	&	2.91	 & 29.5	&N	\\
  &	K0IIb           &	0.001	&	0.005	&	0.15    &	2.8	&	0.02	&	0.10	&0.05	&150	&	250	&	0.67	 & 1.0	&0.00	\\
  &	 + B8IV         &	66.452	&	0.017	&	25.98	&	25.6	&	5.96	&	-0.20	&0.40	&4550	&	11750	&	2.91	 & 29.5	&.02	\\
28&	V415 Car	&	195.300	&	0.000	&	24.29	&	38.6	&	4.41	&	-3.20	&0.00	&4981	&	9388	&	5.99	 & 82.7	&E	\\
  &	G6II      	&	0.000	&	0.010	&	0.10	&	1.0	&	0.02	&	0.20	&0.05	&199.2	&	375.5	&	0.18	 & 2.0	&0.43	\\
  &	 + A1V 	        &	195.300	&	0.000	&	24.29	&	38.6	&	4.41	&	-3.20	&0.00	&4750	&	9650	&	5.99	 & 82.7	&.01	\\
29&	HR 3222	        &	955.130	&	0.327	&	14.06	&	16.85	&	6.03	&	-1.92	&0.00	&4840	&	7000	&	6.78	 & 61.5	&N	\\
  &     K0III           &       0.130	&	0.003	&	0.04	&	0.38	&	0.02	&	 0.10	&0.05	&150	&	300	&	0.45	 & 4.0	&0.18	\\
  &	 + kA8hF2mF4 	&	955.130	&	0.327	&	14.06	&	16.85	&	6.03	&	-1.92	&0.00	&4800	&	7150	&	6.78	 & 61.5	&.02	\\
30&	AL Vel	        &	96.11	&	0.000	&	42.60	&	44.0	&	8.65	&	-1.27	&0.65	&4300	&	11500	&	1.18	 & 85.3	&E	\\
  &     K1II-III        &       0.00	&	0.010	&	0.10	&	1.0	&	0.02	&	0.10	&0.05	&172	&	460	&	0.31	 & 2.0 	&0.71	\\
  &	+ B8V: 	        &	96.11	&	0.000	&	42.60	&	44.0	&	8.65	&	-1.27	&0.55	&4300	&	11100	&	1.00	 & 85.3	&.02	\\
\hline
\end{tabular}\\
\end{table*}

\setcounter{table}{2}
\begin{table*}
\tiny
\setlength{\tabcolsep}{0.075in}
\caption{Observed and modified quantities; continued
\label{tab:3new}}
\indent
\begin{tabular}[]{rlrcrrrrrrrrrcr}
\hline
No.     & Name              &  $P$      & $e$           & $K_1$         & $K_2$         & \V12          &   \DV         & \AV           &\Ta~\,         &\Tb~\,         & plx           &$i$            & Type          \\
        &  spectra          &  day      &               &  km/s         &  km/s         &               &               &               &               &               & mas           &               & GoF           \\
        &                   &           &               &               &               &               &               &               &               &               &               &               &  Z            \\
\hline
31&   RU Cnc        &   10.170  &       0.000   &       70.46  &       67.5   &       10.1    &       -0.30   &0.28    &       4800    &       6400    &       2.64   &  90   &       E       \\
  &   K1IV          &   0.000   &       0.010   &       0.10   &       1.0    &       0.02    &       0.10    &0.05    &       200     &   250         &       0.25   &   2   &       0.59    \\
  &    + F5         &   10.170  &       0.000   &       70.46  &       67.5   &       10.1    &       -0.30   &0.35    &       5000    &       6300    &       2.80   &  90   &       .02     \\
32&   45 Cnc        &   1009.360&       0.461   &       20.04  &       20.75  &       5.62    &       0.15    &0.00    &       5030    &       8500    &       3.52   &  67   &       N       \\
  &   G8III         &   0.120   &       0.002   &       0.06   &       0.23   &       0.02    &       0.10    &0.05    &       150     &   300         &       0.34   &   2   &       0.86    \\
  &    + A3III      &   1009.340&       0.461   &       20.03  &       20.55  &       5.62    &       0.12    &0.00    &       4880    &       9200    &       3.96   &  67   &       .02     \\
33&   $o$ Leo       &   14.498  &       0.000   &       54.80  &       62.08  &       3.52    &       -0.91   &0.00    &       6100    &       7600    &      25.03   &  57.6 &       A       \\
  &   F8IIIm        &   0.000   &       0.000   &       0.08   &       0.16   &       0.02    &       0.10    &0.05    &       200     &   200         &       0.22   &   0.1 &       0.87    \\
  &    + A7m        &   14.498  &       0.000   &       54.75  &       61.95  &       3.52    &       -0.91   &0.04    &       6100    &       7600    &      24.5    &  57.7 &       .02     \\
34&   DQ Leo        &   71.691  &       0.000   &       30.12  &       33.0   &       4.5     &       -0.44   &0.00    &       5300    &       7800    &      14.02   &  50.1 &       A       \\
  &   G7III         &   0.000   &       0.010   &       0.07   &       1.4    &       0.02    &       0.10    &0.05    &       200     &   200         &       0.23   &   0.5 &       1.00    \\
  &    + A7IV       &   71.691  &       0.000   &       30.12  &       33.0   &       4.5     &       -0.44   &0.00    &       5000    &       7700    &      14.65   &  50.1 &       .02     \\
35&   12 Com	    &	396.411	&	0.598	&	24.40  &       30.6   &	      4.80	&	-0.50	&0.00	&	5300	&	8500	&	11.07   &  64	&	 N	\\
  &   G7III         &	0.000	&	0.001	&	0.06	&	0.4	&	0.02	&	0.10	&0.05	&	200	&	500	&	0.24    &   5	&       0.92    \\
  &    + A3IV	    &	396.411	&	0.598	&	24.40	&	30.2	&	4.80	&	-0.70	&0.09	&	5300	&	8700	&	11.6    &  64   &	0.02    \\
36&  3 Boo	    &	36.006	&	0.543	&	52.30	&	59.0	&	5.97	&	-0.07	&0.00	&	5850	&	6750	&	11.15   &  74.5 &        N	\\
  &  G0IV           &	0.000	&	0.002	&	0.19	&	0.6	&	0.02	&	0.10	&0.05	&	150	&	150	&	0.40    &   2.0	&	1.04	\\
  &   + F2p	    &	36.006	&	0.543	&	52.30	&	59.0	&	5.97	&	-0.27	&0.00	&	5550	&	6750	&	11.90   &  74.5	&	0.02	\\
37&  HR 5983	    &	108.206	&	0.000	&	19.83	&	22.41	&	5.79	&	-0.89	&0.06	&	5070	&	9000	&	4.98    &  33	&	N	\\
  &  G7IIIa         &   0.005	&	0.010	&	0.13	&	0.34	&	0.02	&	0.10	&0.05	&	150	&	300	&	0.30	&   2	&	0.45	\\
  &   + A2.5IV      &	108.206	&	0.000	&	19.83	&	22.41	&	5.79	&	-0.89	&0.06	&	5000	&	8600	&	4.70	&  33	&	0.02	\\
38&  HR 6046	    &	2201.00	&	0.68	&	15.51	&	15.69	&	5.63	&	-3.00	&0.00	&	3720	&	4470	&	4.83	&  80	&	A	\\
  &   K3II          &	0.00	&	0.01	&	0.05	&	0.14	&	0.01	&	0.10	&0.05	&	150	&	180	&	.78	&  10	&	0.82	\\
  &   + K0IV	    &	2201.00	&	0.68	&	15.51	&	15.69	&	5.63	&	-3.10	&0.01	&	3720	&	4900	&	4.83	&  80	&	0.02	\\
39&  ASAS-180057    &	269.496	&	0.000	&	35.38	&	35.11	&	10.327	&      -0.037	&1.60	&	4535	&	4211	&	.49     & 88.67	&	E	\\
  &  K4II           &	0.014	&	0.000	&	0.10	&	0.10	&	0.099	&	0.009	&0.10	&	80	&	80	&	.35     &  0.21	&	0.02	\\
  &   + K1II        &	269.496	&	0.000	&	35.38	&	35.11	&	10.327	&      -0.037	&1.60	&	4535	&	4211	&	.467    & 88.67	&	.02	\\
40& ASAS--182510    &	86.650	&	0.000	&	45.12	&	45.45	&	10.87	&	-0.90	&1.35	&	4800	&	4830	&	0.85	& 85.6	&	1.000	\\
  &		    &	0.000	&	0.000	&	0.12	&	0.13	&	0.02	&	0.40	&0.05	&	100	&	107	&	0.25	& 0.8	&	0.74	\\
  &		    &	86.650	&	0.000	&	45.12	&	45.45	&	10.87	&	-0.50	&1.35	&	4650	&	4950	&	0.61	& 85.6	&	1.000	\\
41&  V1980 Sgr      &	40.51	&	0.000	&	42.53	&	41.14	&	10.2	&	-0.30	&0.97	&	4783	&	4600	&	1.31	& 84.2	&	E	\\
  &		    &	0.00	&	0.000	&	0.59	&	0.57	&	0.0	&	0.10	&0.00	&	82	&	163	&	0.2	&  1.4	&	1.19	\\
  &		    &	40.51	&	0.000	&	42.53	&	41.14	&	10.2	&	-0.30	&0.97	&	4487	&	4529	&	1.50	& 84.2	&	.02	\\
42&  V2291 Oph      &	385.0	&	0.311	&	25.30	&	33.1	&	5.64	&	-1.78	&0.60	&	4900	&	11600	&	4.14	& 87	&	 E	\\
  &   G9IIb         &   0.2	&	0.008	&	0.22	&	0.5	&	0.02	&	0.10	&0.05	&	150	&	500	&	0.41	&  2	&	0.35	\\
  &   + B8.5IV      &	385.0	&	0.311	&	25.30	&	33.1	&	5.64	&	-1.78	&0.60	&	4900	&	11100	&	4.34	& 87	&	.02	\\
43&  113 Her	    &	245.325	&	0.101	&	15.48	&	22.58	&	4.57	&	-2.30	&0.00	&	5050	&	9500	&	6.91	& 40.2	&	A	\\
  &  G7II           &	0.006	&	0.005	&	0.09	&	0.30	&	0.02	&	0.10	&0.05	&	150	&	300	&	0.29	&  0.6	&	0.75	\\
  &   + A0V	    &	245.325	&	0.101	&	15.48	&	22.58	&	4.57	&	-2.20	&0.00	&	4850	&	9200	&	7.31	& 39.5	&	.02	\\
44& KIC~10001167    &	120.39	&	0.155	&	25.07	&	26.64	&	10.39	&	-3.60	&0.05	&	4000	&	5160	&	1.23	& 87.6	&	 E	\\
  &		    &	0.00	&	0.002	&	0.10	&	0.85	&	0.09	&	0.09	&0.05	&	99	&	99	&	0.24	& 0.3	&	1.33	\\
  &		    &	120.39	&	0.155	&	25.07	&	26.64	&	10.39	&	-3.70	&0.05	&	4000	&	5160	&	2.20	& 87.6	&	 .02 	\\
45& KIC~5786154     &	197.918	&	0.378	&	24.67	&	25.71	&	14.00	&	-2.38	&0.05	&	4350	&	5800	&	0.305   & 89.1	&	 E	\\
  &		    &	0.001	&	0.001	&	0.02	&	0.08	&	0.09	&	0.09	&0.05	&	99	&	99	&	0.09	& 0.1	&	1.02	\\
  &		    &	197.918	&	0.378	&	24.67	&	25.71	&	14.00	&	-2.38	&0.05	&	4600	&	5600	&	0.305   & 89.1	&	 .02 	\\
46& KIC~3955867     &	33.657	&	0.012	&	37.83	&	45.43	&	14.90	&	-2.50	&0.05	&	4200	&	5700	&	3.15    & 86.75	&	E	\\
  &	 	    &	0.000	&	0.001	&	0.20	&	0.02	&	0.09	&	0.09	&0.09	&	99	&	99	&	0.99    & 0.02	&	1.56	\\
  &	            &	33.657	&	0.012	&	37.83	&	45.43	&	14.90	&	-2.50	&0.05	&	4500	&	5500	&	3.05    & 86.75	&	 .02	\\
47& KIC~7037405	    &	207.108	&	0.228	&	23.56	&	26.02	&	12.00	&	-2.62	&0.05	&	4500	&	6000	&	0.568   & 89.12	&	 E	\\
  &	            &	0.000	&	0.001	&	0.01	&	0.25	&	0.09	&	0.09	&0.05	&	99	&	99	&	0.03	& 0.09	&	1.06	\\
  &		    &	207.108	&	0.228	&	23.56	&	26.02	&	12.00	&	-2.92	&0.05	&	4500	&	6000	&	0.578   & 89.12	&	 .02	\\
48&  9 Cyg	    &	1571.65 &	0.789	&	22.42	&	24.74	&	5.39	&	-0.75	&0.10	&	5050	&	9250	&	5.22	& 117	&	 A	\\
  &  G8IIIa         &	0.38	&	0.002	&	0.12	&	0.33	&	0.02	&	0.10	&0.02	&	200	&	300	&	0.84	&   3	&	0.79	\\
  &   + A2V	    &	1571.65 &	0.789	&	22.42	&	24.94	&	5.39	&	-0.55	&0.10	&	4850	&	9250	&	6.00	& 115	&	.02	\\
49&  SU Cyg	    &	549.2	&	0.343	&	30.07	&	32.2	&	6.98	&	-3.00	&1.10	&	6300	&	8000	&	1.52	&  85	&	 A	\\
  &  F2Iab          &   0.1	&	0.003	&	0.12	&	1.6	&	0.02	&	0.50	&0.20	&	200	&	500	&	0.27	&   2	&	0.71	\\
  &   + (B8 + A0:)  &	549.2	&	0.343	&	30.07	&	32.2	&	6.98	&	-3.00	&1.10	&	6300	&	8000	&	0.92	&  85	&	.01	\\
50&  $\delta$ Sge   &	3705.0  &	0.451	&	7.89	&	8.9	&	3.68	&	-2.50	&0.06	&	3500	&	10500	&	5.49	& 33.5	&	 A	\\
  &  M2IIab         &	3.0	&	0.009	&	0.09	&	2.6	&	0.02	&	0.30	&0.05	&	200	&	500	&	0.72	&  0.3	&	0.70	\\
  &   + B9.5V       &	3705.0  &	0.451	&	7.89	&	8.7	&	3.68	&	-2.00	&0.06	&	3500	&	10900	&	4.60	& 33.5	&	.02	\\
51&  V380 Cyg       &	12.43	&	0.206	&	95.1	&	160.5	&	5.68	&	-3.03	&0.70	&	21750	&	21600	&	.97	& 81.0	&	 E	\\
  &  B1.5III        &	0.00	&	0.010	&	0.3	&	1.2	&	0.02	&	0.05	&0.05	&	280	&	550	&	.02     &  0.5	&	.00	\\
  &   + B2V         &	12.43	&	0.206	&	95.1	&	160.5	&	5.68	&	-3.03	&0.70	&	21750	&	21600	&	.97	& 81.0	&	.02	\\
52&  HD 187669      &	88.387	&	0.000	&	34.444	&	34.458	&	8.88	&	-0.96	&0.38	&	4330	&	4650	&	1.47	& 87.68	&	 E	\\
  &  K2.5III        &   0.001   &	0.001	&	0.015	&	0.015	&	0.01	&	0.05	&0.05	&	70	&	80	&	0.55	&  0.15	&	0.11	\\
  &   + K0-0.5III   &	88.387	&	0.000	&	34.444	&	34.458	&	8.88	&	-0.96	&0.38	&	4330	&	4650	&	1.65	& 87.68	&	 .02	\\
53&  HD 190585      &	171.277	&	0.356	&	33.19	&	33.53	&	9.65	&	-0.15	&2.20	&	4930	&	4930	&	1.27	& 87.05	&	 E	\\
  &		    &	0.001	&	0.001	&	0.05	&	0.05	&	0.09	&	0.09	&0.09	&	199	&	199	&	0.28	&  0.03	&	0.93	\\
  &		    &	171.277	&	0.356	&	33.19	&	33.53	&	9.65	&	-0.15	&2.20	&	4600	&	4600	&	1.77	& 87.05	&	 .02	\\
54&  HD 190361      &	1512.0  &	0.085	&	9.67	&	14.5	&	7.16	&	-1.40	&0.93	&	3800	&	17000	&	0.75	& 33	&	 N	\\
  &  K4Ib           &	1.7	&	0.023	&	0.11	&	1.0	&	0.02	&	0.30	&0.05	&	200	&	500	&	0.36	&  3	&	1.05	\\
  &   + B4IV-V      &	1512.0  &	0.085	&	9.67	&	13.0	&	7.16	&	-1.80	&0.93	&	4000	&	16000	&	1.45	& 33	&	.02	\\
55&  V695 Cyg       &	3784.3  &	0.208	&	13.94	&	24.2	&	3.80	&	-2.59	&0.15	&	3900	&	15500	&	3.69	& 90	&	 E	\\
  &  K4Ib           &	2.0	&	0.009	&	0.19	&	1.0	&	0.02	&	0.20	&0.05	&	200	&	500	&	0.41	&  2	&	2.34	\\
  &   + B5V 	    &	3784.3  &	0.208	&	13.94	&	17.0	&	3.80	&	-2.59	&0.15	&	3800	&	15000	&	3.19	& 90	&	.02	\\
56&  V1488 Cyg      &	1147.51&	0.304	&	16.77	&	34.0	&	3.96	&	-2.00	&0.00	&	3900	&	14000	&	3.08	& 85	&	 E	\\
  &  K5Iab          &	0.00	&	0.010	&	0.10	&	1.0	&	0.02	&	0.30	&0.10	&	156	&	560	&	0.37	& 2	&	2.15	\\
  &   + B7V 	    &	1147.51 &	0.304	&	16.77	&	29.0	&	3.96	&	-3.00	&0.20	&	3900	&	14000	&	3.98	& 85	&	.02	\\
57&  QS Vul	    &	249.18	&	0.011	&	27.10	&	40.0	&	5.18	&	-3.40	&0.18	&	4700	&	12000	&	1.7	& 90	&	 E	\\
  &  G9Ib-II        &	0.10	&	0.008	&	0.21	&	1.0	&	0.02	&	0.30	&0.05	&	200	&	500	&	0.34	&  1	&	1.03	\\
  &   + B8V 	    &	249.18	&	0.011	&	27.1	&	40.0	&	5.18	&	-3.10	&0.18	&	4300	&	12000	&	2.5	& 90	&	.02	\\
58&  $\alpha$ Equ   &	98.810	&	0.000	&	16.53	&	17.9	&	3.92	&	-0.47	&0.00	&	5100	&	8150	&	17.14	& 28.5	&	 A	\\
  &  G7III          &	0.000	&	0.010	&	0.10	&	0.3	&	0.02	&	0.10	&0.05	&	150	&	200	&	0.21	&  1.1	&	1.08	\\
  &   + A4m	    &	98.810	&	0.000	&	16.43	&	18.2	&	3.92	&	-0.47	&0.06	&	5100	&	8200	&	16.54	& 27.0	&	.02	\\
59&  HR 8242	    &	1280.2  &	0.426	&	9.17	&	12.8	&	6.17	&	-1.88	&0.56	&	5210	&	11500	&	1.60	& 29	&	 N	\\
  &  G2Ib           &	0.5	&	0.005	&	0.06	&	1.8	&	0.02	&	0.10	&0.05	&	200	&	500	&	0.42	&  1	&	0.0	\\
  &   + B9IV	    &	1280.2  &	0.426	&	9.17	&	12.8	&	6.17	&	-1.88	&0.56	&	5210	&	11500	&	1.60	& 29	&	.02	\\
60&  HD 208253      &	446.37	&	0.289	&	22.97	&	23.96	&	6.61	&	0.09	&0.12	&	5300	&	9500	&	4.31	& 66.6	&	 N	\\
  &  G7III          &	0.37	&	0.004	&	0.12	&	0.12	&	0.02	&	0.03	&0.05	&	200	&	300	&	0.48	&  2.0	&	0.84	\\
  &   + A2V	    &	446.37	&	0.289	&	22.97	&	23.96	&	6.61	&	0.09	&0.12	&	5050	&	8800	&	3.61	& 66.6	&	.02	\\
\hline
\end{tabular}\\
\end{table*}

We compare the observed data and the computed models in a somewhat unorthodox
way, driven by the facts that
\pn (a) evolutionary tracks are highly non-linear once one moves beyond
the main sequence band,
\pn (b) propagation of errors from the (more or less) directly measured
quantities like $K_2$, $V_{12}$ (the combined apparent visual magnitude),
$\Delta V\equiv V_1-V_2$, or the parallax often gives a
misleading impression of inaccuracy, since many of the errors are correlated,
\pn (c) observational data do not give the {\it initial} masses, period and
eccentricity, which are needed to start the evolutionary code, and
\pn (d) although there will certainly be some mistakes in the theory that goes
into the computed models, such mistakes are inherently {\it systematic} errors,
which cannot be quantified in the way that measurement error can.
\pn What we are mainly looking for is {\it significant} disagreement between
observation and theory, and we feel that a good way to assess the significance
of the disagreement is by using the estimated standard errors of the fundamental data
in a procedure described in Appendix C. This procedure leads to a quantity which
we call Goodness of Fit (`GoF'), which is intended as a crude measure of the
discrepancy between the observational data and our preferred theoretical model
relative to the measurement uncertainties of the observed data. In our collection
of 60 systems we feel that a GoF of less than 1$\sigma$ represents fairly reasonable
agreement, and more than 2.5$\sigma$ represents substantial disagreement.

\def\sc{\scriptstyle}
\def\tpo{$\sc P_0$}
\def\eo{$\sc e_0$}
\def\raf{$\sc R_{1f}$}
\def\rbf{$\sc R_{2f}$}

Table 3 gives three lines per system. The first is the raw observational data, taken from
the literature, and the second is the observational uncertainty from the same source.
The third line is a modified set that we call the `raw theoretical data': a set, but not a
unique set, that fits our preferred theoretical model better. The difference between the
first and third lines, in the sense of an r.m.s. discrepancy normalised by the uncertainties
in the second line, is our Goodness of Fit (GoF) parameter, given at the end of the second
line. Our reasons for adopting this idiosyncratic approach are given in Appendix C.
We believe that if the GoF is less than about 1.5$\sigma$ (in a collection of 60 values), then
the discrepancy between observation and theory is not necessarily serious.

Table 4 also gives three lines per system. The RH half of the first line gives data (masses,
radii etc) derived from the observational data on the first line of Table 3. The LH half
of the second line gives our suggested {\it initial} values of masses, period and eccentricity,
and the RH half gives the consequential current masses, radii etc. We have obviously striven
to ensure that both radii and both temperatures, as well as both masses, are about right. The
RH half of the third line relates to the third line of Table 3 in the same way that the RH half
of the first line relates to the first line of Table 3. The second line of Table 4 also gives
the age of the system (in Myr), and repeats the GoF parameter of Table 3.

If the errors were distributed normally, would expect (in 58 cases, omitting 2 which we
consider to be former triples) 25 with less than 0.5\,$\sigma$, 20 with 0.5 -- 1\,$\sigma$, 10
with 1 -- 1.5\,$\sigma$ and 3 with more than 1.5\,$\sigma$. What we find is 21, 21, 13 and 3
respectively, a considerable degree of consistency. This does not prove that
there is no uncertainty except measurement error; for instance if all discrepancies
were in one direction we should certainly suspect an error in the theory. But it does
mean that we would have to look quite carefully to detect any theoretical error. We
attempt to do just that in \S 5.

The 60 systems which are described in \S4.2 below fall roughly into 4 Classes: \\ (A)
Reasonable agreement (for 42 systems); the agreements range from very good (A+; 12),
through reasonable (A; 16), to rather marginal (A--; 14) but without really significant
disagreement; \\ (B; 15) Often poor agreement that appears to be associated either with
mass loss or the absence of it by stellar
wind from the red-giant component (BM; 11), or with the eccentricity as modified
by tidal friction (BE; 4); these are further subdivided into BM+, BM--, BE+, BE--,
depending on whether the model gave too much or too little of the process; \\
(C; 2) Very poor agreement between the ages of the components, which we suggest
is because the red giant is the merged remnant of a prior {\it sub}-binary; \\
(D; 1) Poor agreement for reasons(s) not yet understood.  \\

For each system, we list below (and in Table~4) what appears to be
the most likely evolutionary state for each component, together with the
Class assignment, A to  D, as explained. A system only qualifies
as A+ if the observational scatter is fairly small {\it and} the theoretical
models (i.e. the circle for the primary, the asterisk for the secondary) agree
well with the mean observed values (squares), as in Figs 1~--~4.

We show online two  sets of three panels for each of the 60 systems. One set of
three panels, like all ten presented here (Figs 1 -- 4), compares the evolutionary
tracks with the derived observational data. The second set compares them with
derived theoretical data, as explained in Appendix C.
\subsection{The binary systems}
\def\e0{$e_0$}
\def\ma{$m_{10}$}
\def\mb{$m_{20}$}
\def\Ra{$\log R_1$}
\def\Rb{$\log R_2$}
\def\Ta{$\log T_1$}
\def\Tb{$\log T_2$}

We abbreviate the main sequence as MS, the Hertzprung Gap as HG, the first giant
branch as FGB, an immediately post-helium-ignition giant as HeIgn, a G--K clump
giant as GKGC, the blue loop as BL, and the asymptotic giant branch as AGB.
\setcounter{table}{3}
\begin{table*}
\setlength{\tabcolsep}{0.050in}
\caption{Comparison of Theory with Observation
\label{tab:4new}}
\indent
\begin{tabular}[]{rlcccccccccrlccc}
\noalign{\smallskip}
    & Name           &       &      &       &       &      & (Myr) &       &      &       &       &       &       &       &       \\
    & Ev. Type       & $P_0$ & $e_0$& \ma   &  \mb  & $n$  &  age  &  $P$  &  $e$ &$m_1$  &$m_2$  & \Ra   & \Rb   &   \Ta & \Tb   \\
    & Quality        &       &      &       &       &      &  GoF  &       &      &       &       &       &       &       &       \\
  1 & SMC-130        &       &      &       &       &      &       & 120.5 & .000 & 1.806 & 1.855 & 1.673 & 1.409 & 3.655 & 3.691 \\
    & AGB + AGB      & 138.8 & .300 & 1.910 & 1.908 & 5066 & 1256. & 119.8 & .000 & 1.848 & 1.856 & 1.673 & 1.369 & 3.638 & 3.678 \\
    & BM--           &       &      &       &       &      & 1.08  & 120.5 & .000 & 1.807 & 1.856 & 1.696 & 1.361 & 3.640 & 3.682 \\
  2 & SMC-126        &       &      &       &       &      &       & 635.0 & .042 & 1.675 & 1.669 & 1.652 & 1.603 & 3.651 & 3.654 \\
    & FGB + FGB      & 593.9 & .100 & 1.725 & 1.724 & 2293 & 1354. & 633.2 & .088 & 1.644 & 1.669 & 1.726 & 1.610 & 3.624 & 3.641 \\
    & A              &       &      &       &       &      & 0.92  & 635.0 & .042 & 1.675 & 1.669 & 1.727 & 1.644 & 3.628 & 3.638 \\
  3 & SMC-101        &       &      &       &       &      &       & 102.9 & .000 & 2.838 & 2.728 & 1.380 & 1.249 & 3.713 & 3.747 \\
    & GKGC + GKGC    & 118.5 & .300 & 2.870 & 2.820 & 2920 & 397.1 & 104.6 & .000 & 2.836 & 2.795 & 1.362 & 1.293 & 3.721 & 3.722 \\
    & A              &       &      &       &       &      & 1.05  & 102.9 & .000 & 2.838 & 2.728 & 1.380 & 1.313 & 3.713 & 3.723 \\
  4 & HD~4615        &       &      &       &       &      &       &302.8  & .435 & 2.818 & 2.518 & 1.547 & 0.603 & 3.643 & 3.940 \\
    & AGB + MS       & 607.3 & .700 & 2.900 & 2.520 & 1773 & 525.0 &303.7  & .422 & 2.797 & 2.520 & 1.513 & 0.580 & 3.651 & 3.930 \\
    & A              &       &      &       &       &      & 0.13  &302.8  & .435 & 2.818 & 2.518 & 1.513 & 0.585 & 3.643 & 3.929 \\
  5 & $\eta$ And     &       &      &       &       &      &       & 115.7 & .003 & 2.391 & 2.259 & 1.028 & 0.933 & 3.703 & 3.699 \\
    & GKGC + FGB     & 133.0 & .300 & 2.368 & 2.268 & 1969 & 809.7 & 117.2 & .000 & 2.327 & 2.264 & 1.040 & 0.912 & 3.698 & 3.705 \\
    & A              &       &      &       &       &      & 0.40  & 115.7 & .003 & 2.371 & 2.260 & 1.041 & 0.920 & 3.699 & 3.703 \\
  6 & SMC-108        &       &      &       &       &      &       & 185.2 & .000 & 4.435 & 4.423 & 1.813 & 1.664 & 3.695 & 3.754 \\
    & BL + BL        & 213.2 & .300 & 4.540 & 4.430 & 3555 & 133.6 & 188.1 & .000 & 4.478 & 4.385 & 1.813 & 1.629 & 3.699 & 3.761 \\
    & A+             &       &      &       &       &      & 0.00  & 185.2 & .000 & 4.435 & 4.423 & 1.813 & 1.664 & 3.695 & 3.754 \\
  7 & BE Psc         &       &      &       &       &      &       & 35.67 & .000 & 1.559 & 1.309 & 1.082 & 0.282 & 3.653 & 3.799 \\
    & FGB + MS       & 38.01 & .300 & 1.630 & 1.380 & 1293 & 2233. & 36.07 & .000 & 1.493 & 1.328 & 1.072 & 0.215 & 3.664 & 3.805 \\
    & A              &       &      &       &       &      & 1.05  & 35.67 & .000 & 1.553 & 1.315 & 1.073 & 0.222 & 3.658 & 3.803 \\
  8 & AS-010538      &       &      &       &       &      &       & 8.07  & .000 & 1.468 & 1.415 & 0.673 & 0.255 & 3.689 & 3.789 \\
    & FGB + MS       & 8.719 & .300 & 1.507 & 1.455 & 1223 & 2869. & 8.43  & .000 & 1.319 & 1.396 & 0.728 & 0.335 & 3.686 & 3.779 \\
    & BM+            &       &      &       &       &      &  0.45 & 8.07  & .000 & 1.468 & 1.415 & 0.727 & 0.335 & 3.689 & 3.788 \\
  9 & AI Phe         &       &      &       &       &      &       & 24.59 & .188 & 1.234 & 1.193 & 0.474 & 0.258 & 3.700 & 3.800 \\
    & FGB + MS       & 23.07 & .250 & 1.290 & 1.246 & 1304 & 4026. & 24.63 & .231 & 1.237 & 1.195 & 0.486 & 0.251 & 3.712 & 3.785 \\
    & A              &       &      &       &       &      &  .64  & 24.59 & .188 & 1.234 & 1.193 & 0.557 & 0.258 & 3.691 & 3.786 \\
 10 & $\tau$ Per     &       &      &       &       &      &       & 1516. & .734 & 2.028 & 1.683 & 1.190 & 0.385 & 3.703 & 3.903 \\
    & GKGC + MS      & 1499. & .739 & 2.180 & 1.748 & 1881 & 1038. & 1514. & .734 & 2.110 & 1.748 & 1.181 & 0.323 & 3.692 & 3.908 \\
    & A              &       &      &       &       &      &  .89  & 1516. & .734 & 2.109 & 1.752 & 1.191 & 0.330 & 3.695 & 3.908 \\
 11 & $\gamma$ Per   &       &      &       &       &      &       & 5328. & .785 & 3.750 & 2.507 & 1.386 & 0.532 & 3.695 & 3.966 \\
    & GKGC + MS      & 5310. & .785 & 3.793 & 2.539 & 1214 & 210.1 & 5306. & .784 & 3.771 & 2.539 & 1.416 & 0.341 & 3.686 & 4.017 \\
    & C              &       &      &       &       &      &   -   & 5328. & .785 & 3.750 & 2.507 & 1.386 & 0.532 & 3.695 & 3.966 \\
 12 & TZ For         &       &      &       &       &      &       & 75.77 & .000 & 2.048 & 1.948 & 0.945 & 0.613 & 3.695 & 3.803 \\
    & GKGC + MS      & 87.28 & .300 & 2.050 & 1.948 & 1420 & 1148. & 75.79 & .000 & 2.039 & 1.946 & 0.949 & 0.536 & 3.688 & 3.817 \\
    & BM+            &       &      &       &       &      &  .57  & 75.77 & .000 & 2.048 & 1.948 & 0.943 & 0.547 & 3.690 & 3.816 \\
 13 & HR 1129        &       &      &       &       &      &       & 6124. & .678 & 4.989 & 4.499 & 1.717 & 0.825 & 3.720 & 4.114 \\
    & GKGC + MS      & 6123. & .679 & 4.880 & 4.460 & 1343 & 111.6 & 6157. & .679 & 4.843 & 4.460 & 1.688 & 0.650 & 3.727 & 4.156 \\
    & A              &       &      &       &       &      &  .80  & 6124. & .678 & 4.883 & 4.462 & 1.687 & 0.688 & 3.720 & 4.146 \\
 14 & OGLE-Cep       &       &      &       &       &      &       & 309.4 & .166 & 4.163 & 4.132 & 1.531 & 1.658 & 3.782 & 3.709 \\
    & BL + GKGC      & 579.5 & .600 & 4.200 & 4.169 & 3089 & 151.2 & 300.9 & .000 & 4.163 & 4.136 & 1.527 & 1.668 & 3.794 & 3.695 \\
    & BE+            &       &      &       &       &      &  .00  & 309.4 & .166 & 4.163 & 4.132 & 1.531 & 1.658 & 3.782 & 3.709 \\
 15 & RZ Eri         &       &      &       &       &      &       & 39.28 & .350 & 1.627 & 1.690 & 0.722 & 0.321 & 3.681 & 3.857 \\
    & FGB + MS       & 39.28 & .350 & 1.927 & 1.690 &  782 & 1287. & 39.51 & .346 & 1.911 & 1.680 & 0.723 & 0.323 & 3.714 & 3.847 \\
    & BM--           &       &      &       &       &      &  .72  & 39.28 & .350 & 1.627 & 1.690 & 0.728 & 0.377 & 3.699 & 3.845 \\
 16 & OGLE-01866     &       &      &       &       &      &       & 251.0 & .241 & 3.576 & 3.577 & 1.675 & 1.442 & 3.657 & 3.726 \\
    & GKGC + FGB     & 856.3 & .750 & 3.600 & 3.580 & 1730 & 202.2 & 255.1 & .156 & 3.585 & 3.577 & 1.644 & 1.497 & 3.672 & 3.689 \\
    & A--            &       &      &       &       &      & 1.01  & 251.0 & .241 & 3.576 & 3.577 & 1.675 & 1.498 & 3.657 & 3.708 \\
 17 & OGLE-03160     &       &      &       &       &      &       & 150.0 & .000 & 1.788 & 1.781 & 1.521 & 1.153 & 3.652 & 3.695 \\
    & FGB + FGB      & 173.0 & .300 & 1.801 & 1.790 & 2040 & 1195. & 149.6 & .000 & 1.762 & 1.786 & 1.530 & 1.153 & 3.653 & 3.692 \\
    & A--            &       &      &       &       &      &  .00  & 150.0 & .000 & 1.788 & 1.781 & 1.521 & 1.153 & 3.652 & 3.695 \\
 18 & $\zeta$ Aur    &       &      &       &       &      &       & 972.2 & .393 & 5.676 & 4.749 & 2.181 & 0.607 & 3.598 & 4.182 \\
    & AGB + MS       & 5595. & .850 & 5.700 & 4.703 & 1637 & 85.76 & 969.2 & .331 & 5.605 & 4.703 & 2.171 & 0.622 & 3.590 & 4.166 \\
    & A              &       &      &       &       &      &  .49  & 972.2 & .393 & 5.591 & 4.712 & 2.171 & 0.605 & 3.598 & 4.173 \\
 19 & OGLE-06575     &       &      &       &       &      &       & 189.8 & .000 & 3.969 & 4.155 & 1.714 & 1.619 & 3.670 & 3.690 \\
    & FGB + FGB      & 218.6 & .300 & 4.180 & 4.160 & 2289 & 147.5 & 186.0 & .000 & 4.149 & 4.133 & 1.680 & 1.706 & 3.679 & 3.673 \\
    & BM--           &       &      &       &       &      &  .00  & 189.8 & .000 & 3.969 & 4.155 & 1.714 & 1.619 & 3.670 & 3.690 \\
 20 & OGLE-EB        &       &      &       &       &      &       & 214.2 & .039 & 3.236 & 3.177 & 1.415 & 1.288 & 3.723 & 3.738 \\
    & GKGC + HG      & 243.9 & .300 & 3.287 & 3.187 & 1778 & 266.7 & 214.0 & .041 & 3.261 & 3.184 & 1.433 & 1.285 & 3.718 & 3.741 \\
    & A+             &       &      &       &       &      &  .00  & 214.2 & .039 & 3.236 & 3.177 & 1.415 & 1.288 & 3.723 & 3.738 \\
\end{tabular}
\end{table*}

(1) SMC-130  (AGB + AGB; BM--): This system illustrates especially well some of the
ambiguities when {\it both} components are highly evolved. The smaller but more massive
giant component could be either on the AGB or else close to the local maximum radius at He
ignition (HeIgn). Although the latter alternative might seem sufficiently short-lived as to be
unlikely, the AGB alternative is not much more long-lived: see panel (c) online.
Fortunately the larger but less massive component can only be on the AGB. We settle for
the AGB + AGB configuration, but this requires the more evolved component to
have lost about 3\% {\it more} of its mass than our model dictates.

(2) SMC-126 (FGB + FGB; A): Since both components appear to be well evolved on the FGB,
they must have started with very nearly equal masses. Our mass-loss algorithm
would have reversed the mass ratio, but only by a small amount. This has not happened;
but the effect is rather slight. Our tidal-friction algorithm did not reduce $e$ from
a hypothetical initial value of 0.1 by more than about 10\%, rather than to the observed
0.042; but this might only mean that the orbit was fairly nearly circular to start
with.

(3) SMC-101 (GKGC + GKGC; A): We obtain acceptable agreement  with both components
in the GKGC, although the theoretical secondary is a little too cool compared
with observation. Our model requires $*2$ to have lost {\it more} mass than $*1$, by
about a factor of two; but this is still a fairly small amount of mass loss.

(4) HD~4615 (AGB + MS; A)  This is neither eclipsing nor interferometric, and
so an inclination of 71\deg 4~was adopted to give a good fit to the theory. The
observational scatter is rather large.
\setcounter{table}{3}
\begin{table*}
\setlength{\tabcolsep}{0.050in}
\caption{Comparison of Theory with Observation, continued
\label{tab:4new}}
\indent
\begin{tabular}[]{rlcccccccccrlccc}
    & Name           &       &      &       &       &      &       &       &      &       &       &       &       &       &       \\
    & Ev. Type       & $P_0$ & $e_0$& \ma   &  \mb  & $n$  & age   &  $P$  &  $e$ &$m_1$  &$m_2$  & \Ra   & \Rb   &   \Ta & \Tb   \\
    & Quality        &       &      &       &       &      & GoF   &       &      &       &       &       &       &       &       \\
\noalign{\smallskip}
 21 & OGLE-09660     &       &      &       &       &      &       & 167.6 & .052 & 2.969 & 2.988 & 1.365 & 1.645 & 3.729 & 3.670 \\
    & GKGC + HeIgn   & 326.0 & .600 & 3.030 & 2.980 & 1944 & 315.7 & 164.7 & .021 & 3.010 & 2.976 & 1.312 & 1.577 & 3.708 & 3.671 \\
    & A-- (BE?)      &       &      &       &       &      & 0.81  & 167.6 & .052 & 2.969 & 2.988 & 1.352 & 1.585 & 3.712 & 3.670 \\
 22 & OGLE-10567     &       &      &       &       &      &       & 117.9 & .000 & 3.347 & 3.184 & 1.405 & 1.558 & 3.705 & 3.672 \\
    & GKGC + HeIgn   & 135.8 & .300 & 3.400 & 3.350 & 2079 & 240.9 & 115.9 & .000 & 3.367 & 3.330 & 1.444 & 1.548 & 3.702 & 3.680 \\
    & BM+            &       &      &       &       &      & 0.0   & 117.9 & .000 & 3.347 & 3.184 & 1.405 & 1.558 & 3.705 & 3.672 \\
 23 & OGLE-26122     &       &      &       &       &      &       & 771.8 & .419 & 3.591 & 3.408 & 1.505 & 1.352 & 3.698 & 3.699 \\
    & GKGC + GKGC    & 773.0 & .420 & 3.600 & 3.450 & 2650 & 252.9 & 765.4 & .402 & 3.538 & 3.426 & 1.489 & 1.352 & 3.699 & 3.701 \\
    & A+             &       &      &       &       &      & 0.0   & 771.8 & .419 & 3.591 & 3.408 & 1.505 & 1.352 & 3.698 & 3.699 \\
 24 & $\alpha$ Aur   &       &      &       &       &      &       & 104.0 & .001 & 2.571 & 2.486 & 1.100 & 0.951 & 3.692 & 3.754 \\
    & GKGC + HG      & 117.7 & .300 & 2.620 & 2.491 & 1758 & 620.3 & 104.5 & .000 & 2.553 & 2.485 & 1.108 & 0.935 & 3.697 & 3.767 \\
    & A+             &       &      &       &       &      &  0.54 & 104.0 & .000 & 2.571 & 2.486 & 1.089 & 0.919 & 3.692 & 3.771 \\
 25 & OGLE-15260     &       &      &       &       &      &       & 157.3 & .000 & 1.427 & 1.440 & 1.621 & 1.355 & 3.635 & 3.673 \\
    & FGB + FGB      & 181.2 & .300 & 1.497 & 1.495 & 2324 & 2043. & 161.5 & .000 & 1.424 & 1.458 & 1.621 & 1.376 & 3.631 & 3.663 \\
    & A              &       &      &       &       &      &  0.0  & 157.3 & .000 & 1.427 & 1.440 & 1.621 & 1.355 & 3.635 & 3.673 \\
 26 & $\delta$ Ori   &       &      &       &       &      &       &  5.73 & .112 & 23.19 & 9.118 & 1.169 & 0.646 & 4.477 & 4.380 \\
    & MS + MS        & 6.147 & .330 & 24.40 & 9.120 &  210 & 6.803 &  5.71 & .109 & 23.25 & 9.121 & 1.175 & 0.621 & 4.483 & 4.375 \\
    & A+             &       &      &       &       &      & 0.25  &  5.73 & .112 & 23.19 & 9.118 & 1.182 & 0.629 & 4.484 & 4.375 \\
 27 & HR 2030        &       &      &       &       &      &       & 66.45 & .017 & 3.926 & 3.984 & 1.563 & 0.712 & 3.658 & 4.070 \\
    & FGB + MS       & 68.60 & .100 & 4.176 & 3.984 &  896 & 157.6 & 65.22 & .034 & 4.168 & 3.984 & 1.561 & 0.699 & 3.663 & 4.076 \\
    & BM--           &       &      &       &       &      & 0.0   & 66.45 & .017 & 3.926 & 3.984 & 1.563 & 0.712 & 3.658 & 4.070 \\
 28 & V415 Car       &       &      &       &       &      &       & 195.3 & .000 & 3.166 & 1.992 & 1.473 & 0.245 & 3.697 & 3.973 \\
    & AGB + MS       & 207.6 & .200 & 3.200 & 2.000 & 1587 & 358.1 & 196.7 & .025 & 3.159 & 2.000 & 1.539 & 0.236 & 3.669 & 3.986 \\
    & A              &       &      &       &       &      & 0.43  & 195.3 & .000 & 3.166 & 1.992 & 1.538 & 0.229 & 3.677 & 3.985 \\
 29 & HR 3222        &       &      &       &       &      &       & 955.1 & .327 & 1.981 & 1.653 & 1.111 & 0.322 & 3.685 & 3.845 \\
    & FGB + MS       & 944.1 & .327 & 1.981 & 1.653 & 1010 & 1204. & 958.5 & .327 & 1.965 & 1.641 & 1.120 & 0.310 & 3.675 & 3.854 \\
    & A+             &       &      &       &       &      & 0.18  & 955.1 & .327 & 1.981 & 1.653 & 1.122 & 0.303 & 3.681 & 3.854 \\
 30 & AL Vel         &       &      &       &       &      &       & 96.11 & .000 & 3.319 & 3.214 & 1.629 & 0.482 & 3.633 & 4.061 \\
    & HeIgn + MS     & 115.2 & .300 & 3.569 & 3.214 & 1095 & 237.1 & 97.66 & .000 & 3.261 & 3.214 & 1.685 & 0.557 & 3.631 & 4.042 \\
    & A--            &       &      &       &       &      & 0.71  & 96.11 & .000 & 3.319 & 3.214 & 1.682 & 0.549 & 3.633 & 4.045 \\
 31 & RU Cnc         &       &      &       &       &      &       & 10.17 & .000 & 1.354 & 1.413 & 0.686 & 0.294 & 3.681 & 3.806 \\
    & FGB + MS       & 11.72 & .300 & 1.550 & 1.425 & 1070 & 2500. & 10.45 & .000 & 1.487 & 1.413 & 0.642 & 0.302 & 3.699 & 3.797 \\
    & BM--           &       &      &       &       &      & 0.59  & 10.17 & .000 & 1.354 & 1.413 & 0.619 & 0.297 & 3.699 & 3.799 \\
 32 & 45Cnc          &       &      &       &       &      &       & 1009. & .461 & 3.235 & 3.124 & 1.294 & 0.808 & 3.702 & 3.929 \\
    & FGB + MS       & 1009. & .461 & 3.171 & 3.091 &  891 & 323.6 & 1012. & .461 & 3.163 & 3.090 & 1.288 & 0.700 & 3.683 & 3.966 \\
    & A--            &       &      &       &       &      & 0.86  & 1009. & .461 & 3.171 & 3.091 & 1.286 & 0.706 & 3.688 & 3.964 \\
 33 & $o$ Leo        &       &      &       &       &      &       & 14.50 & .000 & 2.117 & 1.868 & 0.729 & 0.343 & 3.785 & 3.881 \\
    & HG + MS        &16.70  &.300  &2.110  &1.855  &  592 & 98.68 & 16.76 & .300 & 2.106 & 1.854 & 0.781 & 0.381 & 3.775 & 3.879 \\
    & BE--           &       &      &       &       &      & 0.87  & 14.50 & .000 & 2.099 & 1.855 & 0.746 & 0.360 & 3.785 & 3.881 \\
 34 & DQ Leo         &       &      &       &       &      &       & 71.69 & .000 & 2.163 & 1.974 & 0.905 & 0.439 & 3.724 & 3.892 \\
    & GKGC + MS      & 80.28 &.300  &2.230  &1.974  & 1331 & 908.7 & 72.81 & .000 & 2.086 & 1.973 & 0.981 & 0.435 & 3.696 & 3.884 \\
    & A--            &       &      &       &       &      & 1.00  & 71.69 & .000 & 2.163 & 1.974 & 0.961 & 0.430 & 3.699 & 3.886 \\
 35 & 12 Com         &       &      &       &       &      &       & 396.4 & .598 & 2.696 & 2.150 & 0.952 & 0.404 & 3.724 & 3.929 \\
    & FGB + MS       & 396.4 &.598  &2.630  &2.119  &  706 & 533.0 & 396.9 & .598 & 2.627 & 2.119 & 0.982 & 0.356 & 3.720 & 3.943 \\
    & A--            &       &      &       &       &      & 0.92  & 396.4 & .598 & 2.622 & 2.119 & 0.964 & 0.360 & 3.724 & 3.940 \\
 36 & 3 Boo          &       &      &       &       &      &       & 36.01 & .543 & 1.804 & 1.599 & 0.568 & 0.412 & 3.767 & 3.829 \\
    & HG + MS        &36.01  &.543  &1.820  &1.640  &  761 & 1515. & 36.25 & .535 & 1.795 & 1.515 & 0.644 & 0.362 & 3.736 & 3.828 \\
    & A--            &       &      &       &       &      & 1.04  & 36.01 & .543 & 1.795 & 1.609 & 0.615 & 0.362 & 3.744 & 3.829 \\
 37 & HR 5983        &       &      &       &       &      &       & 108.2 & .000 & 2.775 & 2.455 & 1.197 & 0.479 & 3.705 & 3.954 \\
    & GKGC + MS      &124.6  &.300  &2.825  &2.455  & 1585 & 551.2 & 110.5 & .000 & 2.737 & 2.455 & 1.239 & 0.556 & 3.688 & 3.928 \\
    & A+             &       &      &       &       &      &  0.45 & 108.2 & .000 & 2.775 & 2.455 & 1.241 & 0.531 & 3.699 & 3.934 \\
 38 & HR 6046        &       &      &       &       &      &       & 2201. & .680 & 1.438 & 1.421 & 1.842 & 0.875 & 3.571 & 3.650 \\
    & FGB + FGB      &2722.  &.740  &1.453  &1.437  & 2379 & 3270. & 2206. & .669 & 1.315 & 1.417 & 1.838 & 0.738 & 3.560 & 3.691 \\
    & BM+            &       &      &       &       &      & 0.82  & 2201. & .680 & 1.438 & 1.421 & 1.845 & 0.725 & 3.571 & 3.690 \\
 39 & AS-180057      &       &      &       &       &      &       & 269.5 & .000 & 4.913 & 4.876 & 1.696 & 1.809 & 3.657 & 3.624 \\
    & GKGC + GKGC    & 310.5 & .300 & 5.000 & 4.970 & 2345 & 104.3 & 266.0 & .000 & 4.882 & 4.866 & 1.728 & 1.854 & 3.655 & 3.630 \\
    & A+             &       &      &       &       &      & 0.02  & 269.5 & .000 & 4.913 & 4.876 & 1.714 & 1.828 & 3.657 & 3.624 \\
 40 & AS-182510      &       &      &       &       &      &       & 86.65 & .000 & 3.377 & 3.353 & 1.281 & 1.092 & 3.681 & 3.684 \\
    & FGB + FGB      & 89.34 & .100 &3.380  &3.378  & 1471 & 271.8 & 86.89 & .067 & 3.374 & 3.375 & 1.445 & 1.256 & 3.665 & 3.697 \\
    & BE--           &       &      &       &       &      & 0.74  & 86.65 & .000 & 3.377 & 3.353 & 1.443 & 1.256 & 3.667 & 3.695 \\
\end{tabular}
\end{table*}

\setcounter{table}{3}
\begin{table*}
\setlength{\tabcolsep}{0.050in}
\caption{Comparison of Theory with Observation, continued
\label{tab:4new}}
\indent
\begin{tabular}[]{rlcccccccccrlccc}
    & Name           &       &      &       &       &      &       &       &      &       &       &       &       &       &       \\
    & Ev. Type       & $P_0$ & $e_0$& \ma   &  \mb  & $n$  & age   &  $P$  &  $e$ &$m_1$  &$m_2$  & \Ra   & \Rb   &   \Ta & \Tb   \\
    & Quality        &       &      &       &       &      & GoF   &       &      &       &       &       &       &       &       \\
\noalign{\smallskip}
 41 & V1980 Sgr      &       &      &       &       &      &       & 40.51 & .000 & 1.228 & 1.269 & 1.113 & 1.109 & 3.680 & 3.663 \\
    & FGB + FGB      &35.71  & .300 &1.500  & 1.4995& 2241 & 3064. & 41.09 & .000 & 1.223 & 1.251 & 1.112 & 1.079 & 3.652 & 3.656 \\
    & A              &       &      &       &       &      & 1.19  & 40.51 & .000 & 1.228 & 1.269 & 1.147 & 1.073 & 3.652 & 3.656 \\
 42 & V2291 Oph      &       &      &       &       &      &       & 385.0 & .311 & 3.882 & 2.967 & 1.502 & 0.445 & 3.690 & 4.064 \\
    & GKGC + MS      &1582.  &.800  &3.882  &2.967  & 1344 & 210.7 & 392.0 & .282 & 3.808 & 2.967 & 1.471 & 0.438 & 3.686 & 4.049 \\
    & A+             &       &      &       &       &      & 0.35  & 385.0 & .311 & 3.882 & 2.967 & 1.481 & 0.442 & 3.690 & 4.045 \\
 43 & 113 Her        &       &      &       &       &      &       & 245.3 & .101 & 3.045 & 2.087 & 1.347 & 0.311 & 3.703 & 3.978 \\
    & GKGC + MS      & 262.1 & .230 & 3.195 & 2.181 & 1571 & 390.9 & 249.2 & .109 & 3.148 & 2.181 & 1.375 & 0.327 & 3.679 & 3.963 \\
    & A              &       &      &       &       &      & 0.35  & 245.3 & .101 & 3.181 & 2.181 & 1.375 & 0.322 & 3.686 & 3.964 \\
 44 & KIC1000167     &       &      &       &       &      &       & 120.4 & .155 & 0.859 & 0.808 & 1.329 &  0.212& 3.602 & 3.713 \\
    & FGB + MS       & 117.9 & .200 & 1.050 & 0.830 & 1196 & 11824.& 126.9 & .147 & 0.966 & 0.812 & 1.109 & -0.094& 3.643 & 3.715 \\
    & BM--           &       &      &       &       &      & 1.33  & 120.4 & .155 & 0.859 & 0.808 & 1.077 & -0.061& 3.602 & 3.713 \\
 45 & KIC5786154     &       &      &       &       &      &       & 197.9 & .378 & 1.062 & 1.019 & 1.044 & 0.189 & 3.638 & 3.763 \\
    & FGB + MS       & 178.8 & .390 & 1.140 & 1.090 & 2595 & 8712. & 201.6 & .384 & 1.059 & 1.029 & 0.956 & 0.180 & 3.661 & 3.760 \\
    & A+             &       &      &       &       &      & 1.02  & 197.9 & .378 & 1.062 & 1.019 & 0.957 & 0.227 & 3.663 & 3.748 \\
 46 & KIC3955867     &       &      &       &       &      &       & 33.66 & .012 & 1.103 & 0.919 & 0.912 &-0.008 & 3.623 & 3.756 \\
    & FGB + MS       & 33.41 & .300 & 1.260 & 0.955 & 1544 & 5845. & 33.68 & .001 & 1.112 & 0.919 & 0.850 &-0.049 & 3.671 & 3.744 \\
    & A--            &       &      &       &       &      & 1.56  & 33.66 & .012 & 1.103 & 0.919 & 0.817 &-0.010 & 3.653 & 3.740 \\
 47 & KIC7037405     &       &      &       &       &      &       & 207.1 & .228 & 1.267 & 1.148 & 1.125 & 0.239 & 3.653 & 3.778 \\
    & FGB + MS       & 185.5 & .232 & 1.350 & 1.210 & 1793 & 4594. & 206.8 & .226 & 1.266 & 1.148 & 1.124 & 0.151 & 3.653 & 3.778 \\
    & A              &       &      &       &       &      & 1.06  & 207.1 & .228 & 1.267 & 1.148 & 1.121 & 0.176 & 3.653 & 3.778 \\
 48 & 9 Cyg          &       &      &       &       &      &       & 1572. & .789 & 2.938 & 2.662 & 1.261 & 0.550 & 3.703 & 3.966 \\
    & GKGC + MS      & 1917. & .820 & 2.843 & 2.552 & 1172 & 437.1 & 1593. & .793 & 2.800 & 2.552 & 1.234 & 0.497 & 3.681 & 3.969 \\
    & A              &       &      &       &       &      &  0.79 & 1572. & .789 & 2.838 & 2.551 & 1.241 & 0.515 & 3.686 & 3.969 \\
 49 & SU Cyg         &       &      &       &       &      &       & 549.2 & .343 & 5.957 & 5.563 & 1.505 &       & 3.799 &       \\
    & BL + (MS + MS) & 1239. & .700 & 6.050 & 5.550 & 1451 & 687.0 & 453.5 & .000 & 6.000 & 5.550 & 1.733 &       & 3.798 &       \\
    & BE+            &       &      &       &       &      &  0.71 & 549.2 & .343 & 5.957 & 5.563 & 1.724 &       & 3.799 &       \\
 50 & $\delta$ Sge   &       &      &       &       &      &       & 3705. & .451 & 4.073 & 3.611 & 2.362 & 0.520 & 3.544 & 4.021 \\
    & AGB + MS       & 3705. & .451 & 3.987 & 3.525 & 2001 & 213.1 & 3705. & .451 & 3.882 & 3.525 & 2.348 & 0.660 & 3.540 & 4.041 \\
    & A              &       &      &       &       &      &  0.70 & 3705. & .451 & 3.887 & 3.525 & 2.428 & 0.671 & 3.544 & 4.037 \\
 51 & V380 Cyg       &       &      &       &       &      &       & 12.43 & .206 & 13.13 & 7.782 & 1.218 & 0.614 & 4.337 & 4.334 \\
    & MS + MS        & 12.00 & .206 & 13.46 & 7.782 &  281 & 16.95 & 12.60 & .208 & 13.12 & 7.781 & 1.217 & 0.628 & 4.331 & 4.330 \\
    & A+             &       &      &       &       &      &  0.00 & 12.43 & .206 & 13.13 & 7.782 & 1.218 & 0.614 & 4.337 & 4.334 \\
 52 & HD 187669      &       &      &       &       &      &       & 88.39 & .000 & 1.502 & 1.501 & 1.407 & 1.105 & 3.636 & 3.667 \\
    & FGB + FGB      & 101.8 & .300 & 1.560 & 1.555 & 2269 & 266.1 & 95.54 & .000 & 1.456 & 1.489 & 1.356 & 1.025 & 3.634 & 3.670 \\
    & BM+            &       &      &       &       &      &  0.11 & 88.39 & .000 & 1.502 & 1.501 & 1.356 & 1.054 & 3.636 & 3.667 \\
 53 & HD 190585      &       &      &       &       &      &       & 171.3 & .356 & 2.170 & 2.148 & 1.426 & 1.396 & 3.693 & 3.693 \\
    & FGB + FGB      & 369.7 & .700 & 2.180 & 2.1794 & 1849& 914.2 & 144.6 & .233 & 2.140 & 2.155 & 1.426 & 1.379 & 3.645 & 3.650 \\
    & A--            &       &      &       &       &      &  0.93 & 171.3 & .356 & 2.170 & 2.148 & 1.380 & 1.350 & 3.663 & 3.663 \\
 54 & HD 190361      &       &      &       &       &      &       & 1512. & .085 & 8.125 & 5.419 & 2.434 & 0.862 & 3.580 & 4.230 \\
    & GKGC + MS      & 1494. & .085 & 6.436 & 4.767 & 1139 & 58.19 & 1484. & .072 & 6.399 & 4.767 & 2.047 & 0.536 & 3.618 & 4.194 \\
    & A--            &       &      &       &       &      & 1.05  & 1512. & .085 & 6.409 & 4.767 & 2.062 & 0.532 & 3.602 & 4.204 \\
 55 & V695 Cyg       &       &      &       &       &      &       & 3784. & .208 & 12.92 & 7.441 & 2.234 & 0.522 & 3.591 & 4.190 \\
    & AGB? + MS?     & 3785. & .208 & 6.000 & 4.900 & 1792 & 76.12 & 3836. & .208 & 5.922 & 4.900 & 2.357 & 0.624 & 3.563 & 4.180 \\
    & D              &       &      &       &       &      &  2.34 & 3784. & .208 & 5.972 & 4.897 & 2.358 & 0.599 & 3.580 & 4.176 \\
 56 & V1488 Cyg      &       &      &       &       &      &       & 1148. & .304 & 9.113 & 4.495 & 2.237 & 0.683 & 3.591 & 4.146 \\
    & GKGC + MS      &3444.  &.750  & 6.300 & 3.650 & 1058 & 60.14 & 1113. & .303 & 6.268 & 3.650 & 2.188 & 0.407 & 3.591 & 4.132 \\
    & A--            &       &      &       &       &      &  2.14 & 1148. & .304 & 6.317 & 3.653 & 2.185 & 0.431 & 3.591 & 4.146 \\
 57 & QS Vul         &       &      &       &       &      &       & 249.2 & .011 & 4.649 & 3.150 & 1.984 & 0.531 & 3.672 & 4.079 \\
    & AGB + MS       &287.0  & .300 & 4.739 & 3.150 & 1653 & 135.1 & 240.5 & .000 & 4.590 & 3.150 & 1.951 & 0.409 & 3.612 & 4.079 \\
    & A--            &       &      &       &       &      &  1.03 & 249.2 & .011 & 4.649 & 3.150 & 1.950 & 0.421 & 3.633 & 4.079 \\
 58 & $\alpha$ Equ   &       &      &       &       &      &       & 98.81 & .000 & 2.000 & 1.847 & 0.986 & 0.426 & 3.708 & 3.911 \\
    & GKGC + MS      & 112.1 & .300 & 2.408 & 2.156 & 1319 & 703.8 & 98.60 & .000 & 2.360 & 2.156 & 1.013 & 0.457 & 3.698 & 3.913 \\
    & A--            &       &      &       &       &      &  1.10 & 98.81 & .000 & 2.388 & 2.156 & 1.013 & 0.449 & 3.708 & 3.914 \\
 59 & HR 8242        &       &      &       &       &      &       & 1280. & .426 & 5.326 & 3.816 & 1.722 & 0.729 & 3.717 & 4.061 \\
    & GKGC + MS      & 1459. & .500 & 5.325 & 3.815 & 1183 & 90.88 & 1256. & .415 & 5.281 & 3.815 & 1.789 & 0.467 & 3.656 & 4.133 \\
    & C              &       &      &       &       &      &   --  & 1280. & .426 & 5.326 & 3.816 & 1.722 & 0.729 & 3.717 & 4.061 \\
 60 & HD 208253      &       &      &       &       &      &       & 446.3 & .289 & 2.770 & 2.656 & 0.970 & 0.474 & 3.724 & 3.978 \\
    & GKGC + MS      & 449.8 & .300 & 2.800 & 2.656 & 1274 & 468.5 & 451.3 & .293 & 2.758 & 2.656 & 1.117 & 0.619 & 3.697 & 3.936 \\
    & A              &       &      &       &       &      &  0.84 & 446.4 & .289 & 2.770 & 2.656 & 1.110 & 0.595 & 3.703 & 3.944 \\
\end{tabular}
\end{table*}

(5) $\eta$~And (GKGC + FGB; A): A reasonable fit was obtained, but the
scatter in masses was considerable. The near-circularity of the orbit suggests
that at least the primary has evolved past its local maximum radius at helium
ignition. The secondary is arguably too small to be in the GKGC too.

(6) SMC-108 (BL + BL; A+): We get good agreement, with the more massive and
larger component near the end of the Blue Loop and the less massive and smaller just
starting the BL. We expect both stars to have lost 1 or 2\% per cent of their mass.

(7) BE~Psc (FGB + MS; A): This is an RS CVn binary. In view of the small
scatter, one might hope for a better fit, but we believe the one shown is
acceptable.  Almost all systems were started with a default
rotational period of 2\,d for each component. In this relatively low-mass
system, that leads to some modest dynamo-driven wind mass loss close to the
main sequence in both components, until the rotation is slowed to $\sim$\,5 d
(Table~2). The orbit is later circularized by tidal friction. The
model predicts rather more mass loss than the observations suggest. The parameters
of this system resemble those of RZ Eri, yet the effects of both
tidal friction and mass loss are very different.

(8) AS-010538 (FGB + MS; BM+): Our model gives a dynamo-driven wind
that causes $*1$ to lose about 4 times as much mass as it apparently has.
In this it is similar to, but milder than, BE Psc, and dissimilar to RZ Eri and RU Cnc.

(9) AI~Phe (FGB + MS; A): We started $e$ at 0.25 -- somewhat on the low side
for an unevolved system of this period -- in order to have it reduced by tidal
friction to roughly its present value.
We used $Z$ = 0.01, following Andersen et al.~(1988).  By combining
dynamo-driven wind and Roche-lobe overflow, the model makes the primary evolve
to a white dwarf of 0.32\,\Msun, while $M_2$ increases to 1.35\,\Msun~and $P$ to 124\,d.

(10) $\tau$~Per (GKGC + MS; A): Both masses are quite uncertain, with
$K_2$ = 23.0~$\pm$~4.0\,\kms, but a $K_2$ of 25 gives a good agreement,
using an appropriately sub-solar $Z = 0.01$.  There was reasonable agreement in
the $\log T, \log L$ plane only if the parallax was modified from 12.8
to 13.5\,mas (2\,$\sigma$), but other quantities agreed better and the overall
GoF was 0.9\,$\sigma$.

(11) $\gamma$~Per (GKGC + MS; C): The observations are unusually exact, and
our models cannot give a good agreement (Fig~4); $*2$ is much too evolved for
its low mass relative to $*1$. We suggest that $*1$ is the remnant of a
relatively recent merger of a sub-binary with a period of a few days (see \S5.1).

(12) TZ~For (GKGC + MS; BM+): Because the system is somewhat metal-rich, we used
an opacity table for $Z$ = 0.03. The fit seems reasonable, but only because the
theoretical mass-loss rate was reduced, for this system only, by a factor of 20.
Tidal friction circularized the orbit during the helium-ignition phase. If the
primary were on the First Giant Branch, the orbit would not have circularized.

(13) HR~1129 (GKGC + MS; A): Reasonable agreement is achieved if $Z$ = 0.01, but
with considerable observational scatter; modest modifications to several
variables at the 0.5\,$\sigma$ level make the agreement good.

(14) OGLE-Cep (BL + GKGC; BE+): This system is also in the LMC.  The fit appears
to be good, but our theoretical tidal friction has reduced $e$ to zero prematurely.

(15) RZ~Eri (FGB + MS; BM--): Although there is some uncertainty in the
primary mass, it is clear that the primary must have lost substantial mass,
about 20 times as much mass as our dynamo-driven wind model predicts.

(16) OGLE-01866 (GKGC + FGB; A--): The theoretical secondary is rather too cool and faint,
but not unacceptably so. We estimate that $*1$ is just beyond He ignition, and $*2$ just
before. The model tidal friction may be a little too strong, but not by much.

(17) OGLE-03160 (FGB + FGB; A--): The masses must have been fairly closely equal on the
ZAMS, and are now still closely equal a long way up the FGB. Our theoretical primary,
approaching its Roche lobe, has lost rather more mass than observation suggests, but still
only a few per cent.

(18) $\zeta$~Aur (AGB + MS; A): The agreement is good, but with considerable
observational scatter largely because the RV of the hot star (and thence $q$)
is intrinsically difficult to measure. We had to start from a high, but not
unreasonably high, eccentricity (0.85) to have it reduced to something like
the currently measured eccentrity.

(19) OGLE-06575 (GKGC + GKGC; BM--): The large near-equal radii suggest that the initial
masses must have been closely equal, but the larger star is now less massive by $\sim$5\%.
Our mass-loss model does not give that much.

(20) OGLE-EB (GKGC + HG; A+): Because the system is in the LMC, we used an
opacity table for $Z$ = 0.004. Both components appear to be hotter than the
giant branch, which might argue for lower $Z$ still. But the fit appears to be
very good.

(21) OGLE-09660  (GKGC + HeIgn; A--): The primary has reached the local minimum
of radius near the start of the GKGC, while the secondary is close to the local
maximum (HeIgn) preceding it. Our model suggests that slightly more mass has been
lost by the primary than by the secondary, which is not in strong conflict with
the observations but not strongly supported either.

(22) OGLE-10567 (GKGC + HeIgn; BM+): Our models fit best if we suppose that the currently
larger star was initially the less massive, and is still approaching the GKGC after
He ignition, while the slightly smaller but initially more massive star has already reached it.

(23) OGLE-26122 (GKGC + GKGC; A+): A very good fit. The primary is near the end of
the GKGC, and the secondary near the beginning. The model suggests a rather slight
amount of mass loss from both components, which is neither strongly supported nor
strongly contradicted by the observations.

(24) $\alpha$~Aur (GKGC + HG; A+): Good agreement is obtained, and with an
apparently rather precise set of observed data.  We have already commented on a
substantial discrepancy between two apparently accurate tabulations of
parameters for this system, and we must await a more detailed understanding of
the {\it systematic} errors that must be present. However this does not mean
that {\it every} observational data set has similar problems. $\alpha$~Aur is
remarkable in that (a) both components are giants with rather similar
temperatures and spectra, and (b) the hotter component is rotating at least 10
times faster than the cooler, so its lines are unusually broad and shallow. In
order to allow $*2$ to rotate as rapidly as it does (in 8 days), we had to
start the components at zero age with a rotation period of about 1.05 days,
whereas 2 days was the normal starting value.

(25) OGLE-15260 (FGB + FGB; A): Both components are well up the FGB. The larger component
is the less massive, presumably because of mass loss as it approaches its Roche lobe. Our
model gives quite good agreement with this.

(26) $\delta$ Ori (MS + MS; A+) The primary is an O9.5 bright giant, but we expect it
to be in the MS band because if it were in the HG it would be evolving measurably
on a timescale of 100 yrs. The evolutionary tracks look somewhat complicated, but
only because the evolution was followed up to and including RLOF, and through
two brief contact phases to the reversal of the mass ratio.

(27) HR 2030 (FGB + MS; BM--): Although the scatter is considerable, particularly
in the parallax, the observations appear to favour substantial mass loss from
$*1$, at roughly 20 times more than our tentative dynamo-driven wind model predicts.

(28) V415 Car (AGB + MS; A): The observations are relatively tightly
constrained, but to make them agree better with the model we modified the temperatures
by about 1\,$\sigma$. The {\it average} modification was about 0.43\,$\sigma$.

(29) HR 3222 (FGB + MS; A+): The observational scatter in mass is substantial
but the agreement with theory is good. We can wonder
whether $*1$ is {\it beyond} helium ignition, which would involve a degenerate
flash at this relatively low mass; the orbit is wide enough to allow this. The
`Horizontal branch' for this model is a
very short stub against the giant branch. We suggest it is just a coincidence
that the radius of $*1$ is in fact fairly close to the radius expected on the
horizontal branch, alias the GK giant clump.

(30) AL Vel (HeIgn + MS; A--)  The giant must be either close to helium ignition
or on the AGB. In fact evolution
immediately subsequent to helium ignition is not as rapid as one might expect --
about 3.10$^5$\,yr -- so we favour that. The observational scatter prevents a
more firm conclusion. The amount of dynamo-driven wind appears to be quite
substantial (about 6\% of $M_1$), but also appears to be about right.

(31) RU Cnc (FGB + MS; BM--): This system is near the bottom end of the range of
periods that we preselected for our sample.  It is a well-known RS CVn system,
and though there is some scatter it seems rather clear that $*1$ has less mass
than $*2$, and has presumably experienced substantial dynamo-driven wind --
perhaps about 5 times as much as our model gives.

(32) 45 Cnc (FGB + MS; A--): The period is sufficiently long that the absence of
circularity is not a problem.  Dynamo-driven wind appears to be only a minor
effect.  The Li abundance in the giant is more consistent with a giant on the
first giant branch than in the GK giant clump. Modifications of $\sim$
0.5\,$\sigma$ were made to three measured quantities (the temperatures and
parallax).

(33) o Leo (HG + MS; BE--): This is one of five systems where one component
is in the Hertzsprung gap. Both components have metallic-line characteristics,
so although the more evolved one has a rapidly growing a convective envelope
which contains $\sim$3.10$^{-4}$\Msun~according to the model, it cannot yet be
deep enough to mix the composition back to normality; the diffusive separation
process must therefore have previously extended to a depth somewhat in excess
of this amount of mass. The agreement is very good, and the scatter is small.
However, the tidal friction model is unsatisfactory: it fails to circularise the
orbit until $*1$ is about 15\% larger.  This seems to be an indication that the
theoretical tidal friction should be increased, but there are other systems,
e.g. SU~Cyg, which present the opposite case (\S5.3).

(34) DQ Leo (GKGC + MS; A--): This seems to be a fairly solid case of
post-helium-ignition structure. The orbit was circularised during helium
ignition when $*1$ was 3 times larger.  The modest amount of mass loss
predicted by the dynamo-driven wind model appears to be acceptable.

(35) 12 Com (FGB + MS; A--): The radius of $*1$ is more compatible with that of
a giant on the first giant branch than in the GK giant clump, but not by
much. The substantial eccentricity also favours that solution, though not
conclusively; so does the detection of Li in its spectrum.

(36) 3 Boo (HG + MS; A--): An acceptable fit is achieved near the local minimum
radius towards the red end of the Hertzsprung gap.  However, the observed
difference in magnitudes is rather less than the model requires, by about
2\,$\sigma$. The overall GoF is about 1.04$\sigma$.

(37) HR~5983 (GKGC + MS; A+): There is substantial scatter in mass, but the
likely GK giant clump state is consistent with the orbit having been
circularized at helium ignition. Slightly modified temperatures, at the level
of $\sim$1\,$\sigma$, gave better agreement in the H--R diagram. The overall
fit is 0.45\,$\sigma$.

(38) HR~6046 (FGB + FGB; BM+): This was a difficult system to model. The
modelled mass loss is mainly by single red-giant wind rather than dynamo-driven
wind, and perhaps the single red-giant wind is an overestimate even although it
does not seem unreasonable that a $\sim$1.5\,\Msun star should reduce its mass
to $\sim$1\Msun~on the first giant branch (and then further to a white dwarf
mass of $\sim$0.8\,\Msun~on the asymptotic giant branch). Note the oddity that
the horizontal branch is to the {\it red} side of the first giant branch; this
arises because the giant branch for a 1\,\Msun~star is substantially to the right
of that for the original 1.45-\Msun-star. Note also that because of the almost
equal masses the first giant branch of $*2$ (green) is so much on top of that
for $*1$ that a portion of the latter is hidden.

(39) AS-180057 (GKGC + GKGC; A+): Our model suggests that the slightly less massive
component  was originally the slghtly  more massive one.

(40) AS-182510 (FGB + FGB; BE--): There is particularly large scatter in the
parallax, and so in the luminosities and radii; so although the fit does not seem
good it can be considered marginally acceptable. The theoretical orbit does not
circularise until $*1$ is about 20\% larger.

(41) V1980 Sgr (FGB + FGB; A): Both components lie slightly on the hot side of
the model tracks, which might indicate a lower metallicity than the solar value
that was used. The masses are so nearly equal that the evolutionary tracks lie
largely on top of each other.

(42) V2291~Oph (GKGC + MS; A+): Here we found relatively small scatter, and a
good fit. The modelled tidal friction appears to be slightly too strong, even
though we assumed an initial $e$ of $\sim$0.8.

(43) 113~Her (GKGC + MS; A): A good fit was obtained, after modifications
averaging $\sim$0.75\,$\sigma$ to the temperatures, parallax and inclination
were applied. The fact that the modelled temperatures were both slightly too
low might be an indication that the metallicity is slightly sub-solar.

(44) KIC~10001167 (FGB + MS; BM--): Our theoretical FGB  is too hot, by
$\sim$ 500K; our $*2$ is about right. We have to suppose that $*1$ was initially
more massive than at present, in order to get the evolutionary age down to something
which is just believable (12 Gyr), but still our model does not get down to the
observed 0.859\Msun; starting at 1.05\Msun it decreases to 0.97\Msun.

(45) KIC~5786154 (FGB + MS; A+): A very good fit, with the FGB primary having lost
rather more mass than the MS secondary. The low masses require an age of about 8 Gyr.

(46) KIC~3955867 (FGB + MS; A--): Very hard to reconcile the theoretical temperatures
with those observed. The temperature of $*2$, at 0.92\Msun, is observed to be 5700K,
more appropriate to our near-ZAMS models of $\sim$ 1.0\Msun; and our theoretical red
giant at about the observed radius is about 500K hotter than observed.

(47) KIC~7037405 (FGB + MS; A): The theoretical $*2$ is slightly too small and faint.

(48) 9~Cyg (GKGC + MS; A): There is substantial scatter in mass, but the fit
was reasonable.

(49) SU~Cyg (BL + $\left[{\rm MS\ +\ MS}\right]$; BE+): Triple systems were not to
be included in the sample, but this is one of the very few Cepheids with a
reasonably well-determined mass \citep{Evans.Bolton.1990}, and it is fairly
crucial in modelling overshooting (see \S3.1). The modelled $*2$ was a fictional
entity with the combined mass of the sub-binary.

(50) $\delta$~Sge (AGB + MS; A): The observations show substantial scatter, and
our model gives an acceptable fit. The scatter of plusses for $M_1$ looks one-sided,
but that is because several plusses are beyond the panel to the right.

(51) V380~Cyg (MS + MS; A+): This system was included to demonstrate the need
to have {\it considerable} overshooting at high masses.  If overshooting
were not enhanced substantially by a factor of $\sim$3 over what appears to
prevail at 2--6\Msun, the track would pass well below the observed point (see
\S3.1).

(52) HD~187669 (FGB + FGB; BM+): The model gave too much mass loss, but otherwise
fits well. The orbit was circularised in the nick of time.

(53) HD~190585 (FGB + FGB; A--): The initial masses must have been very closely
equal for the radii to differ by less than 10\% high up on on the FGB. The
theoretical tidal friction is perhaps too strong, but only by a small amount.

(54) HD~190361 (GKGC + MS; A--): There is substantial scatter, but the fit is
acceptable. The model tidal friction was sufficiently weak that we had to start
with $e$ as low as 0.12, rather small perhaps for an orbit of this size.

(55) V695~Cyg (AGB? + MS?; D): The observational data in the first line of the
corresponding entry in Table~3 leads to masses of 13.5 and
8.1\,\Msun. These were not used, because no theoretical model will
give $*2$ as massive as 8\,\Msun~and at the same time as faint as $\log L \sim$
3.0 and as cool as $\log T \sim$ 4.16.  In contrast, for V380~Cyg a good fit to
the same mass for $*2$ was obtained for with $\log L \sim$ 3.5 and $\log T \sim$
4.33.  $K_2$ was therefore modified from 24~$\pm$~1\,\kms~to 17\,\kms
(3rd line of Table~3).  The magnitude difference and the parallax were also
modified in order to achieve the fit shown. The system resembles quite closely
both HD~190361 and V1488~Cyg, which give marginally acceptable fits. We classify
this as a complete misfit (Class D, our only one), and do not for the present
offer an explanation.

(56) V1488~Cyg (GKGC + MS; A--): The fit is not at all good, but may just be
tolerable. Substantial modification was needed to $K_2$, the parallax and the
difference in magnitudes, averaging 2.15\,$\sigma$ overall.

(57) QS Vul (AGB + MS; A--): There is not much observational scatter, but substantial
modification was needed to the parallax and the difference in magnitudes,
averaging 1.0\,$\sigma$~overall

(58) $\alpha$~Equ (GKGC + MS; A--): An acceptable fit, but with substantial scatter.

(59) HR~8242 (GKGC + MS; C): Considerable scatter was present, but even so $*2$
is considerably `over-evolved', as was found for $\gamma$~Per.

(60) HD~208253 (GKGC + MS; A): A reasonable fit was obtained, but with
substantial scatter. The hot component has enhanced Zr and Ba: the latter is often
associated with s-processing that occurred in a companion that was once on the AGB
and is now a white dwarf. It does not seem impossible that there is such a companion
in the present system, perhaps with a period of 10 -- 15 yrs.

\section{Discussion}

Three of the 60 cases are not at all well approximated by our models; two of
those ($\gamma$~Per and HR~8242) were classed as `C', meaning that the
secondary is considerably more evolved than it should be, and one was classed
as `D', meaning that there was no reasonable possibility of a fit.
We discuss these below.

\subsection{The two class-C systems}

We suggest what we believe is a likely explanation for the two C systems.  The
same explanation can also be applied to two other, possibly related, systems
which are not included in our analysis: OW~Gem \citep{Griffin.Duquennoy.1993} and
V643~Ori \citep{Imbert.1987}. In this context we also consider the remarkable
triple system V453~Cep \citep[HD~216572;][]{GriffinR.R.2009}, which is
particularly relevant to the discussion.

V453 Cep consists of a K giant and a sub-binary of two late B stars.  The {\it
outer} orbit is only 55\,d, the inner orbit is 1.2\,d, and the masses are
2.65+(2.6+2.5)\,\Msun.  Although it would be difficult to display all three
components on the kind of H--R diagrams of Figs 1 -- 4, the data given by
\citet{GriffinR.R.2009} indicate that it is a markedly `under-evolved' system,
with both B dwarfs being very near the ZAMS while the K giant, of only very
slightly more mass, is obviously quite highly evolved.

This system, and other triples like HR~6497 \citep{GriffinR.R.2012} pose the
question: why is it only the {\it secondaries} that are
sub-binaries?  At first glance the answer is obvious: because if the primary
had once been a comparably small sub-binary, it would have suffered some severe
interaction by now as it grew to giant dimensions. However, the primary did not
`know' that when it began its evolution.  Therefore, if 6 out of $\sim$46
systems had sub-binary secondaries, one could reasonably expect that another 6
would have had sub-binary {\it primaries}, though such sub-binaries would have
changed dramatically as a result of evolution.  Those starting with nearly
equal masses in the sub-binary would probably have evolved to a semi-detached
Algol-like configuration -- quite like Algol itself, which is a triple with an
F dwarf in a 2-yr outer orbit. Other sub-binaries which started from more
unequal masses may have merged by now into a single star, and that is precisely
the kind of evolution that could lead to an apparently `under-evolved' secondary
such as that in $\gamma$~Per.

One can suggest three main possible outcomes: \\ (1) If the mass ratio $q_1$
($\equiv M_{11}/M_{12}$) is in the range 1--1.4 the system becomes an Algol,
and evolves to longer period even although magnetic braking and tidal friction
will remove some angular momentum. \\ (2) If $q_1\sim$1.4--2 the stars come
rapidly into contact, and form a contact binary that evolves mainly by
magnetically wind-driven angular momentum loss
to larger $q$, either decreasing its period or at least not increasing
it by much.  After slow evolution, on a timescale of perhaps
10$^8$--10$^9$\,yrs, and at a large $q$ which might be in the range 10 -- 25,
it merges into a single star. \\ (3) If $q_1 > 2$ the system merges rather
quickly.

Nelson \& Eggleton (2001) considered the conservative evolution of a large
number of mostly close binaries, and found that while those with a mass ratio
in the range 1 to 1.5 were usually able to evolve into classical semidetached
systems, those with mass ratios above this tended to evolve into a rather
catastrophic regime of mass transfer. The exact boundary between mild and
catastrophic mass transfer is very unclear, and it probably depends on both the
total mass and the initial orbital period.  Let us suppose a mass ratio of 1.4
is fairly critical.  If we have a triple with masses of (say) (2.5 + (2.6 +
1.2)), the 2.6-\Msun~component will evolve first, interact rather
catastrophically with its 1.2-\Msun~companion, possibly just after the
2.6-\Msun~component has left the main sequence. The result of this catastrophic
interaction may be a merger, leaving a red giant with mass 3.8\,\Msun, or a bit
less if some mass is lost in the process. This red giant will be left with a
relatively widely orbiting companion of 2.5\,\Msun, but the important point is
that this companion will be quite substantially evolved since for most of its
life it was only 4\% less massive than the component that became a red
giant. This situation could (we believe) be just what we now see in $\gamma$~Per
and HR~8242.

That binaries can merge and become single stars was demonstrated
extraordinarily well by the remarkable observations of V1309~Sco
\citep{Tylenda.al.2011}.  They found that this star was an eclipsing binary
until 2008; it then underwent a 10 mag.~eruption lasting about 2\,yr, and is
now a single star. Such an event is just what we hypothesize happened to the
primary of $\gamma$~Per perhaps 10\,Myr ago; the age of the secondary is about
450\,Myr, if we interpret its position on its evolutionary track as being
consistent with that model.

Statistics of the frequency of different levels of multiplicity are not
especially definitive, but \citet{Eggleton.Tok.2008} estimated values for the
complete sample of 4559 systems brighter than Hipparcos magnitude 6.00. The
frequencies of multiplicities 1, 2, \dots, 7 were found to be 2718, 1437, 285,
86, 20, 11 and 2 (planets being excluded).  Undoubtedly more are still to be
found, although this is a sample that has mostly been studied quite intensively
for over 200 years, probably more thoroughly studied from the point of view of
multiplicity than any other sample of comparable size. That sample is certainly
not representative of stars in general, since (for instance) it includes hardly
any M dwarfs, except as secondaries, although M dwarfs are much the commonest
type of star in the Galaxy. But it may be fairly representative of those stars
over about 1\,\Msun, which are certainly the ones most likely to have evolved
significantly in the lifetime of the Galaxy and therefore the ancestors of
systems that now contain a red giant. The above statistics say that among the
$\sim$1800 systems that are at least binary, $\sim$400, or 22\%, are at least
triple. Thus it should not be particularly surprising if among 60 binaries 3 or
4 were once triples containing a rather close sub-binary. In fact the
statistics are a little more compelling than that since (as already mentioned)
at least 6 sub-binaries have been found among 46 systems classified in the first
instance as composite-spectrum binaries. This latter statistic excludes those
triples which happen to be in {\it wide} hierarchical systems; in fact though
we have excluded almost all systems with sub-binaries from Table~1, several systems in
that Table have {\it wide} companions that we do not discuss.

Two eclipsing systems that perhaps should be in Table~1 but are not, for
reasons we now describe, are OW~Gem \citep[F2Ib--II + G8IIb; 1259\,d; 6 +
4\,\Msun;][]{Griffin.Duquennoy.1993} and V643~Ori \citep[K2III + K7III;
54.2\,d, 3.3 + 1.9\,\Msun;][]{Imbert.1987}. Because both components, in both
systems, are evolved giants, it would be reasonable to expect that their masses
would be nearly equal, but that is clearly not the case by a wide margin; we
cannot even say for sure which is the more evolved and might therefore qualify
as the primary.  It is possible that OW Gem is a {\it former} triple (Eggleton
2002), but the same explanation might seem less likely for V643~Ori, given its
short period. However Griffin \& Griffin (2009) have found that V453~Cep
(HD~216572) is a triple with an {\it outer} period of 54.7\,d, and so it does
now appear a little more tenable.  The initial parameters in the
two cases may have been something like [(4.1 + 1.9\Msun; 4\,d) + 4\,\Msun;
1259\,d] and [(1.95 + 1.35\,\Msun, 3\,d) + 1.9\,\Msun, 55\,d].

  Although it is only tangentially relevant to the 60 systems mainly discussed
here, V453 Cen is interesting on its own, since the two members of the sub-binary
are apparently much less evolved than the third body, which is only slightly
more massive.  Griffin \& Griffin (2009) show that the components are markedly
non-co\ae val, but in the opposite sense to $\gamma$ Per -- the MS dwarfs are `under-evolved'
relative to the giant. So our explanation for $\gamma$ Per will obviously not work in
this case. We wonder if it is an example of a process described by Pflamm-Altenburg
and Kroupa (2007): a binary forms in one young star-forming region, is ejected
as stars evidently are from its natal cluster, and travels in the Galaxy for
some considerable time before colliding with another much younger star-forming
region. Gravitational focussing makes the effective cross-section of the second
cluster considerably larger than the cross-section of just another star or binary.
Then there is a gravitational encounter between the original binary and a much
younger binary in the second cluster, which can lead to a bound triple and
an ejected component. Something rather similar, though not exactly similar, was
suggested by Gualandris et al.~(2004) as explaining the fact that the two close
components of $\iota$ Ori appear to be of markedly different ages, although both
ages (7 and 3.5 Myr) are arguably consistent with different parts of the Orion
Nebula star-forming region.

\subsection{BM systems}
Many of our theoretical systems lost appreciable amounts of mass, in the range
of 1 -- 10\%, due to the
mass-loss processes described in \S3.2. In several cases there was reasonable
agreement with the observations, but in 11 cases there was a marked discrepancy:
AS-010538, TZ For, OGLE-10567, HR 6046 and HD 187669 were modelled with too much
mass loss, by factors of about  3 -- 20; and SMC-130, RZ~Eri, OGLE-06575, HR 2030,
RU Cnc and KIC 10001167 were modelled with too {\it little} mass loss by similar
factors.

A particularly significant pair of cases, we think, is shown in the top and bottom
panels of Fig. 2: RZ~Eri and BE~Psc. These systems have rather similar masses
and period, (1.63 + 1.69\Msun, 39d) and (1.49 + 1.33\Msun, 36d), and seem
very probably to be in the same evolutionary state (FGB + MS). Yet the first
seems to have had its primary mass reduced by about 20\%, while in the second
there is no clear evidence of any mass loss at all, although our model suggests
abut 5\% which is nevertheless larger than the observations compel. A curious
and suprising further difference is that BE Psc has circularised its orbit, while
RZ Eri is still far from circular. In fact our Tidal Friction model is consistent
with this, a little surprisingly, but one might expect that tidal friction that
had circularised the orbit in such binaries would speed up the rotation of the
giant, and thus cause it to be more rather than less mass-lossy.

Another interesting pair are $\alpha$ Aur (Fig. 1) and TZ For (online only). It
can be seen that the first is fitted well with our regular mass-loss model: indeed
we seem to need just the mass loss that the model dictates to get the best fit. But
for TZ For (and for that system only) we reduced our Dynamo-Driven Wind model by
a factor of 15 in order to get comparably good agreement.

Other discrepant systems do not appear in quite such illustrative pairs, but
nevertheless give either too much or too little mass loss by factors up to about
twenty. There does
not seem to be a single `normalisation' factor that could reduce those
disparities. The wide scatter suggests to us that there is something inherently
chaotic in the process of mass loss driven by dynamo activity, and ultimately
by rapid rotation.  There is certainly something chaotic, at least in the loose
sense, about solar activity, which underwent a marked decrease in the 17$^{\rm
th}$ Century, the `Maunder Minimum' (Eddy  1976) which lasted for about 70 years.
The equations of magnetohydrodynamics, which no doubt govern activity, are
certainly complex enough to allow chaotic solutions, and it could be that some
stars get into a very long-lived, more-or-less permanent Maunder-like minimum,
while others get into very long-lived active states.

This might seem a rather strong conclusion to draw from a rather limited set of
data, and from a dynamo-driven wind model which is at best only sketchy, and yet
there is no denying the individuality which is manifested by nominally similar
systems like  BE~Psc and RZ~Eri. It would be difficult to devise a continuous
formulation that was sophisticated enough to encompass both systems, but it
would not be difficult for an inherently chaotic process to produce two very
different outcomes from rather similar initial circumstances.

\subsection{BE systems}

Twenty five of our systems, being started with eccentricities larger than their
current values, had them reduced substantially and satisfactorily to about the current value by
our tidal friction model. A further 12 were sufficiently wide that their current
eccentricity, which is substantially non-zero, could reasonably be assumed to have
been unchanged. But 2 systems were classed as BE+, meaning that the model tidal
friction was too strong to allow the present non-zero eccentricities to be
maintained; and 2 were classed as BE--, meaning that it was too weak to explain
the present zero eccentricities. The two former -- both Cepheids, perhaps
coincidentally -- were at longish periods (309d, 549d) and the two latter at
shortish periods (87d, 15d). So it might be that our model depends too steeply
on period or separation, but the evidence is not compelling.

      As a long shot, we wonder whether the interaction of Cepheid pulsations
with tidal effects might actually {\it create} eccentricity which was
previously damped out. This might account for the two BE+ systems.

\subsection{The Class-C and Class-D systems}

For the two class-C systems, we did not evaluate a GoF
value since (a) it was already obvious that both secondaries were considerably
`over-evolved', and (b) we could identify what seemed like a very probable
explanation. Of the 58 remaining systems only one (V695 Cyg, with GoF 2.3$\sigma$)
seemed so aberrant that we are not able to offer an explanation for it.

\section{Conclusions}

Approximately 42 of the sample of 60 binaries can be fitted reasonably well
by theoretical models, provided core convective overshooting is modified to
allow rather more overshooting for masses at and above $\sim$13\,\Msun~than for
models below $\sim$6\,\Msun. A further 11 are not well fitted mainly (it appears)
because mass loss seems to be a very erratic, perhaps chaotic, process that
will be difficult to describe with a single formula. A further 4 are discrepant
in eccentricity, presumably because of inadequacy in the tidal friction algorithm.
A further 2 have `over-evolved secondaries', which can be attributed rather well
to the possibility that they were originally triple but have experienced a merger
of a former close sub-binary. The remaining 1 is harder to explain, but we should
not rule out the possibility of observational error at the 2\,$\sigma$ level or above.

      We are deeply indebted to Dr R. E. M. Griffin for much helpful discussion
regarding many of these systems, and their presentation. KY gratefully acknowledges the
support provided by the Turkish Scientiﬁc and Technical Research Council (T\"UB\.ITAK-113F097 and 111T270).

\newpage
\section {APPENDICES
}
\subsection{APPENDIX A: Calculation of core overshooting}
\def\o{\over}
The code uses variables $r^2$ and $\mu \equiv m^{2/3}$, instead of $r$ and $m$,
because these variables vary linearly with each other and with log\,$P$,
log\,$T$ and log\,$\rho$ at and near the centre, and so allow a central
difference approximation to be used down to and including the central
meshpoint. They also allow one to define a characteristic `central mass
parameter' $\mu_c$, thus:

$$P= P_0\left[1-a_1{\mu\o\mu_c}+a_2\left({\mu\o\mu_c}\right)^2+\dots\right]\ , $$
$$\mu_c = {3P_0\o 2G\rho_0^{4/3}},\ \  a_1={3\o 4}\left({4\pi\o
3}\right)^{1/3}.\eqno(A1)$$

The choice of the numerical factor $3/2$ in $\mu_c$ is arbitrary (provided
$a_1$ is adjusted correspondingly), but this choice means that $M_c \equiv
\mu_c^{3/2}$ is roughly the mass of the star if it is on the ZAMS, to $\pm$ 7\%
over the range 1--100\,\Msun; but it also gives the $\it helium$ core mass
correctly if the star has evolved to contain a homogeneous centrally-convective
helium-burning core.  The method supposes that if the extent of core convective
overshooting varies from star to star it is likely to be determined by the mass
of the {\it star} if the star is H-rich in its core, but by the mass of the
{\it He core} if the core is H-exhausted.  \par The overshooting model is a
modification, $\dOS$, to the usual Schwarzschild convection criterion $\nabla_r
> \nabla_a$, thus:
$$\nabla_r > \nabla_a - \dOS .\eqno(A2)$$

In the absence of any detailed numerical or physical modelling, we adopt $\dOS$
as
$$\dOS = {C_{\rm OV} \o 2.5+20\beta\,' +16\beta\,'\,^2} ,\eqno(A3)$$

\noindent with $\beta$\,$'$ the ratio of radiation pressure to gas pressure,
and $C_{\rm OV}$ given by

$${C_{\rm OV,A}+(C_{\rm OV,B}-C_{\rm OV,A})\max\left[0,\min\left(1,
{\mu_c-\mu_{c,\rm A}\o \mu_{c,\rm B}-\mu{c,\rm_A}}\right)\right]} .\eqno(A4)$$
The $\beta$\,$'$ term is to compensate for the fact that at large masses both
$\nabla_r$ and $\nabla_a$ tend to the value 4/3, the former because Thomson
scattering tends to dominate over Kramers' opacity, and the latter because
radiation pressure tends to dominate over gas pressure.

The coefficients of Equn (A4) are
$$C_{\rm OV,A} = 0.11,\  C_{\rm OV,B} = 0.30 , \eqno(A5)$$
$$\mu_{c, \rm A}^{3/2} \equiv M_{c, \rm A} = 6.3 \,M_{\odot},
~~~\mu_{c,\rm B}^{3/2} \equiv M_{c,\rm B} = 14.8\,M_{\odot} . \eqno(A6)$$
$C_{OV}$ is thereby chosen to be fairly small for stars less than $\sim$6\,\Msun,
including SU Cyg, but to increase substantially until about 14\,\Msun~in order
to model V380~Cyg.  A further substantial increase seems unlikely, but can
hardly be ruled out from the models presented here.

\subsection{APPENDIX B: Converting from Basic to Derived Observational Data}

We attempt to represent the measured quantities and their uncertainties in a
homogeneous way in Table~3, as follows. For each system either 10 or 11
directly observed basic quantities are listed; the 11th, the inclination of the
orbit to the line of sight, is measured for eclipsing or astrometric systems,
but not (except by inference from fitting to theoretical models) for 10 systems
which are neither. Inferred values of $i$ are indicated by an $N$ in the last
column.

The 11 quantities are: \def\DV{\Delta V} $$P,\ e,\ K_1,\ K_2,\ V_{12},\ \DV,\
A_V,\ T_1,\ T_2,\ 1/d\ , i, \eqno(B1)$$
i.e. period (days), eccentricity, RV
amplitudes (\kms), combined (Johnson) $V$ magnitude, difference in magnitude
($V_1-V_2$), IS absorption, temperatures, parallax, i.e. reciprocal distance,
from Hipparcos (van Leeuwen 2007) or Gaia \citep{gaia1,gaia2}, and orbital inclination. In principle,
these should determine the following 8 derived quantities which are convenient
to plot, and to compare with theory: $$ M_1,\ M_2,\ \log R_1,\ \log R_2,\ \log L_1,\ \log L_2,\ \log T_1,\
\log T_2.\eqno(B2)$$

The transition from (B1) to (B2) requires certain standard formul\ae\ given
below, plus (i) a table of bolometric corrections as a function of temperature (we
have used Flower 1996), (ii) the effective temperature of the Sun (5771.8\,K) and the
bolometric magnitude of the Sun (4.7554), from Mamajek (2015), and (iii) a constant,
$1.03614907.10^{-7}$, that relates RV amplitude to mass (Taylor \& Weisberg 1989).
In quite a number of published papers, even rather recent ones, an earlier value
of that constant, ($1.0385.10^{-7}$) has been used. Eclipsing double-lined
binaries often yield a distance $d$ which is independent of a direct parallax
measurement, and may in some cases be more accurate than measured parallaxes;
if (for instance) an orbital period is close to a year it is not easy to
disentangle orbital motion from parallactic motion.

\def\o{\over}

In terms of the 11 observed `raw' quantities (B1), we get the `derived'
quantities (B2) by way of the formul\ae

$$M_1=1.03614907.10^{-7}{(1-e^2)^{3/2}(K_1+K_2)^2K_2P\o \sin^3i}\eqno(B3)$$
$$M_2={M_1K_1\o K_2}$$ $$M_{V12}=V_{12}-A_V+5\,\log(10/d)\eqno(B4) $$
$$M_{V,1}=M_{V12}+2.5\,\log(1+10^{0.4\DV})$$ $$M_{V,2}=M_{V,1}-\DV\eqno(B5)$$
$$M_{{\rm Bol},i}=M_{V,i}+BC\,(T_i),\ i=1,2\eqno(B6)$$
$$\log L_i=-0.4(M_{{\rm Bol},i}-4.7554),\ i=1,2\eqno(B7)$$
$$R_i^2\,=L_i\,\left({5771.8\o T_i}\right)^4,\ i=1,2 .\eqno(B8)$$

\subsection{APPENDIX C: Determining a `Goodness of Fit' Parameter}

We start from the 11 directly measured quantities, as identified in (B1) and
listed, for each system, in the first line of each three-line entry for each
of the 60 systems in Table~3. We refer to this set as the `raw observational
data'. Secondly, we use the algorithms of Appendix B to turn them into 8 values
that we call the `derived observational data' such as masses and radii, as
identified in (B2). These are given for each system towards the right in the
first line for each system in Table~4.  Thirdly, we estimate largely by
intuition what {\it initial} values of masses, period and eccentricity, as
given in the left half of the second line in Table 4, will lead through evolution
to something like the observed present masses, etc. The stellar evolution code then
leads to values in the right-hand half of the second line of Table 4, which we call
the `derived theoretical data'. In an ideal world they would be exactly the same
as the `derived observational data' above them. But the world is not ideal and
so the two half-lines differ.  Fourthly, we determine some `raw theoretical data',
i.e. we attempt to reverse the process that led from (B1) to (B2). This process
is of course not unique, but we try to make selections that give the closest
approximation of the `raw theoretical data' to the `raw observational data'. The
raw theoretical data are listed in the third line for each system in Table~3.
Then finally we are in a position to make a direct comparison using the measurement
uncertainties which are found in the observations and which are listed in the second line for each
system in Table~3. The r.m.s difference between the third line and the first line,
as normalised by the second line, is then our `Goodness of Fit' or GoF parameter.

    Well before we attempted to fit all 60 systems it became clear that many
were not going to agree with our mass-loss recipes. If it had seemed likely that
this disagreement could be eliminated by using a different mass-loss recipe, we
would have tried that, e.g. by scaling the mass-loss rates by some empirical
factor. But the chaotic nature of the disagreement made that pointless. Consequently,
in attempting to match the derived theoretical data to the derived observational
data we normally concentrated on just 5 values: $M_2,\ \log R_1,\ \log R_2,\
\log T_1$\ and $\log T_2$; obviously the luminosities will be right if the
temperatures and radii are. Thus our GoF parameter may be quite good even if $M_1$
\ is quite bad. We label such systems as Class BM.

\end{document}